%% file: ANA-HDBS-2018-53-PAPER.tex
\documentclass[PAPER, cernpreprint, orcidlogo, atlasdraft=false, texlive=2023, UKenglish, texmf, orcidlogo]{atlasdoc}
 
\usepackage[subfigure]{atlaspackage}
\usepackage{atlasbiblatex}

\usepackage{dcolumn}
\usepackage{multirow}
\usepackage{enumerate}
\usepackage{amsmath}
\usepackage{atlasphysics}
 
\graphicspath{{logos/}{figures/}{SlicePlots_Individual/}}

\AtlasTitle{Searches for exclusive Higgs and $Z$ boson decays into a vector quarkonium state and a photon using $139$~fb$^{-1}$ of ATLAS $\sqrt{s}=13$~TeV proton--proton collision data}
 
\AtlasAbstract{
Searches for the exclusive decays of Higgs and $Z$ bosons into
a vector quarkonium state and a photon are performed in the  $\mu^+\mu^-\,\gamma$ final state with a proton--proton collision data sample corresponding to an integrated luminosity of
$139$~fb$^{-1}$ collected at $\sqrt{s}=13$~TeV with the ATLAS detector at
the CERN Large Hadron Collider.
The observed data are compatible with the expected backgrounds.
The 95\% confidence-level upper limits on the branching fractions of the
Higgs boson decays into $J/\psi\,\gamma$, $\psi(2S)\,\gamma$, and $\Upsilon(1S,2S,3S)\,\gamma$ are found to be
$2.0\times10^{-4}$, $10.5\times10^{-4}$, and $(2.5,4.2,3.4)\times10^{-4}$, respectively, assuming Standard Model production of the Higgs boson.
The corresponding 95\% CL upper limits on the
branching fractions of the $Z$ boson decays are $1.2\times10^{-6}$, $2.4\times10^{-6}$, and $(1.1,1.3,2.4)\times10^{-6}$.
An observed 95\% CL interval of $(-133,175)$ is obtained for the $\kappa_c/\kappa_\gamma$ ratio of Higgs boson coupling modifiers, and a 95\% CL interval of
$(-37,40)$ is obtained for $\kappa_b/\kappa_\gamma$.
}
 
\author{The ATLAS Collaboration}
 
\AtlasRefCode{HDBS-2018-53}

\PreprintIdNumber{CERN-EP-2022-128}
\AtlasJournal{EPJC}
 
\AtlasJournalRef{Eur. Phys. J. C 83 (2023) 781}
\AtlasDOI{10.1140/epjc/s10052-023-11869-1}

\hypersetup{pdftitle={ATLAS document},pdfauthor={The ATLAS Collaboration}}
 
\newcommand\progname[1]{{\scshape #1}}
\def\ee{\ensuremath{e^+ e^-}}

\def\epem{\ensuremath{e^+ e^-}}
\def\mumu{\ensuremath{\mathrm{\mu^+ \mu^-}}}
\def\mm{\ensuremath{\mathrm{\mu\mu}}}
\def\tautau{\ensuremath{\mathrm{\tau^+ \tau^-}}}

\def\leplep{\ensuremath{\ell^+ \ell^-}}
\def\lnu{\ensuremath{\ell \nu}}
\def\ptmumu{\ensuremath{p_{\textrm{T}}^{\,\mm}}}

\def\ptmu{\ensuremath{p_{\textrm{T}}^{\,\mu}}}
\def\ptmumu{\ensuremath{p_{\textrm{T}}^{\mm}}}
\def\etamu{\ensuremath{\eta^{\,\mu}}}

\def\ptgamma{\ensuremath{p_{\textrm{T}}^{\gamma}}}
\def\etagamma{\ensuremath{\eta^{\gamma}}}

\def\HtoJpsiGamma{\ensuremath{H\to\Jpsi\,\gamma}}
\def\HtoPsiGamma{\ensuremath{H\to\psi(2S)\,\gamma}}
 
\newcolumntype{d}[1]{D{.}{\pm}{#1}}
\newcolumntype{h}[1]{D{.}{}{#1}}

\newcommand{\tabscript}[3]{
\setlength{\fboxrule}{0pt}
\fbox{\ensuremath{#1^{#2}_{#3}}}
 
}
\begin{document}
 
\maketitle

\section{Introduction}
\label{sec:intro}
 
The Higgs boson, $H$, was first observed by the ATLAS~\cite{Aad:2008zzm} and CMS~\cite{CMS-TDR-08-001} collaborations in 2012~\cite{HIGG-2012-27,Chatrchyan:2012ia} with a mass of approximately $125$~\GeV.
Detailed measurements of its properties~\cite{ATLAS:2022vkf,CMS:2022dwd} have confirmed its role in the
spontaneous breaking of electroweak symmetry and the mass generation of the massive vector bosons~\cite{Englert:1964et,Higgs:1964ia}.
In the Standard Model (SM), the mass generation for fermions is implemented through Yukawa interactions.
The ATLAS and CMS collaborations have reported observations of Higgs boson decays into a pair of $\tau$-leptons~\cite{HIGG-2019-09,CMS-HIG-19-010},
a pair of bottom quarks~\cite{HIGG-2018-04,CMS-HIG-18-016},
and associated production of Higgs bosons with top-quark pairs~\cite{HIGG-2018-13,CMS-HIG-17-035}.
These measurements represent a complete observation of Higgs boson couplings to third-generation charged fermions, and are in agreement with SM expectations.
Recently, evidence has been reported for the Higgs boson coupling to muons in the second generation through the decay $H\to\mumu$~\cite{CMS-HIG-19-006,HIGG-2019-14}.
Direct searches for $H\to\ccbar$ have been performed by both the ATLAS and CMS collaborations~\cite{HIGG-2017-01,ATLAS:2022ers,CMS-HIG-18-031,CMS:2022psv}, as have searches for $H\to\ee$ decays~\cite{HIGG-2018-58,CMS-HIG-13-007}, but no further experimental evidence currently exists for the Higgs boson couplings to the first and second generations of fermions.
Searches for potential beyond-the-SM (BSM) couplings of the Higgs boson have also been performed by the ATLAS and CMS collaborations,
including searches for flavour-changing neutral currents via the $t$-quark decays  $t\rightarrow cH$ and $t\rightarrow uH$~\cite{HIGG-2016-26,TOPQ-2017-15,CMS-TOP-19-002,CMS-TOP-20-007},
and the lepton-flavour-violating decays $H\to e\mu$, $H\to e\tau$ and $H\to \mu\tau$~\cite{HIGG-2018-58,ATLAS:2019pmk,CMS-HIG-20-009}.
No evidence for these couplings has been found.
 
\begin{figure}[!htbp]
\centering
\subfigure[Direct amplitude]{\includegraphics[width=0.33\linewidth]{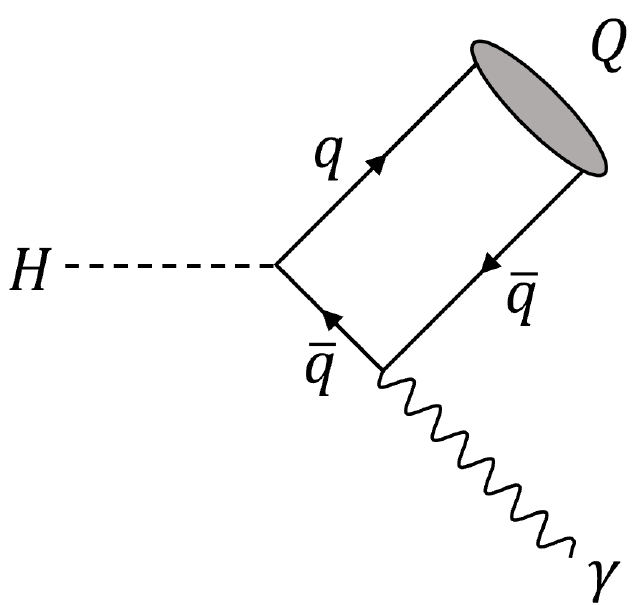}\label{fig:directAmp}}\hspace{2.5cm}
\subfigure[Indirect amplitude]{\includegraphics[width=0.33\linewidth]{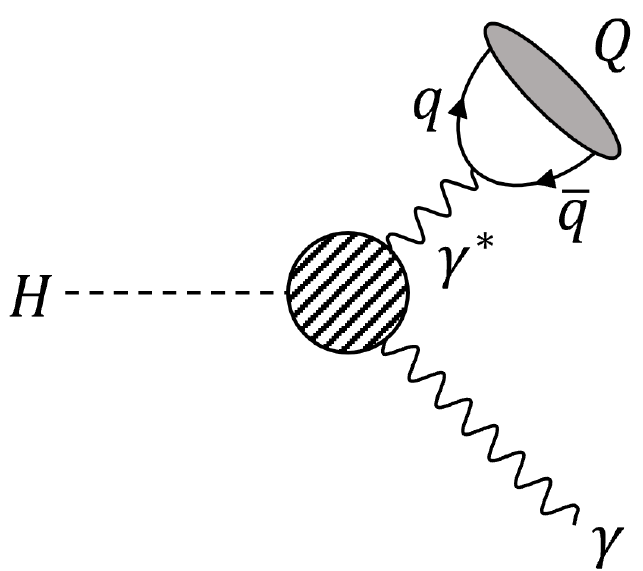}\label{fig:indirectAmp}}
\caption{Feynman diagrams depicting the~\subref{fig:directAmp} direct amplitude and~\subref{fig:indirectAmp} indirect amplitude contributing to the $H\rightarrow\mathcal{Q}\,\gamma$ process, where $\mathcal{Q}$ is a vector quarkonium state.
The hatched circle in~\subref{fig:indirectAmp} denotes a set of one-loop diagrams.\label{fig:higgsFeynman}}
\end{figure}

A direct method for accessing the couplings of the first- and second-generation quarks is the study of inclusive $H\rightarrow q\bar{q}$ decays.
While these channels have large branching fractions, their experimental sensitivity is substantially obscured by large multi-jet backgrounds.
Searches for the exclusive, radiative decays of the Higgs boson into a vector meson state and a photon offer an alternative way to probe the quark Yukawa couplings~\cite{HBranching,VatoHBranching,Koenig:2015pha}.
Although their branching fractions are comparatively small, these radiative decays have a distinct experimental signature, which helps suppress the large multi-jet backgrounds that affect the $H\rightarrow q\bar{q}$ searches.
Figure~\ref{fig:higgsFeynman} shows Feynman diagrams depicting the $H\rightarrow\mathcal{Q}\,\gamma$ process, where $\mathcal{Q}$ is a vector quarkonium state.
There are two primary contributions to the decay amplitude: the direct amplitude ${\cal A}_\text{dir}$ occurs via the quark Yukawa coupling;
the indirect amplitude ${\cal A}_\text{ind}$ occurs at the one-loop level in the SM through the $H\rightarrow\gamma\gamma^*$ decay, where the virtual photon subsequently fragments into a vector quarkonium state.
The two processes interfere destructively, and despite being loop-induced, the indirect amplitude is typically the more dominant contribution in decays of the Higgs boson into a vector meson state and a photon.
 
Higgs boson decays in the charmonium sector, $\HtoJpsiGamma$ and $\HtoPsiGamma$, offer an
opportunity to access both the magnitude and the sign of the charm-quark Yukawa coupling~\cite{HBranching,VatoHBranching};
the corresponding decays in the bottomonium sector, $H\to\Upsilon(1S,2S,3S)\,\gamma$,
can provide information about the real and imaginary parts of the bottom-quark coupling to the Higgs boson~\cite{Koenig:2015pha}.
Studies of these decays complement searches for the inclusive $H\rightarrow c\bar{c}$ and $H\rightarrow b\bar{b}$ decays.
The results of recent independent calculations of the branching fractions expected for these decays in the SM are presented in Table~\ref{tab:HQgammaSM}, and are of the order of $10^{-6}$ for $\HtoJpsiGamma$
and of order $10^{-9}$ to $10^{-8}$ for $H\to \Upsilon(1S,2S,3S)\,\gamma$~\cite{Bodwin:2014bpa,Bodwin:2016edd,Bodwin:2017wdu,Koenig:2015pha,Zhou:2016sot,Brambilla:2019fmu}.
The branching fraction for $\HtoPsiGamma$ is expected to be $(1.03\pm 0.06)\times 10^{-6}$.
This was obtained via a private communication from the authors of Ref.~\cite{Bodwin:2017wdu}, who used an estimate of the value of the order-$\nu^2$ non-relativistic QCD long-distance matrix element, where $\nu$ is the velocity of the heavy quarks in the $\mathcal{Q}$ rest frame.
It is noted that the branching fractions for decays in the bottomonium sector are small compared to those in the charmonium sector:
in this case there is an almost perfect cancellation between the direct and indirect decay amplitudes, caused by the mass of the $b$-quark being large compared to the masses of quarks in the first and second generations.
 
\begin{table}[!htbp]
\begin{center}
\caption{Recent calculations of the $H\rightarrow\mathcal{Q}\,\gamma$ branching fractions expected in the Standard Model.\label{tab:HQgammaSM}}
\begin{tabular}{|c|c|c|c|}
\hline
Vector            & \multicolumn{3}{c|}{SM branching fraction, $\mathcal{B}\left(H\rightarrow\mathcal{Q}\,\gamma\right)$} \\ \cline{2-4}
quarkonium state  & Ref.~\cite{Koenig:2015pha} (2015)               & Refs.~\cite{Bodwin:2016edd,Bodwin:2017wdu} (2017)               & Ref.~\cite{Brambilla:2019fmu} (2019) \\ \hline
$\Jpsi$           & $\tabscript{2.95}{+0.17}{-0.17}\times 10^{-6}$ & $\tabscript{2.99}{+0.16}{-0.15}\times 10^{-6}$ & $\tabscript{3.01}{+0.15}{-0.15}\times 10^{-6}$ \\
$\Upsilon(1S)$    & $\tabscript{4.61}{+1.76}{-1.23}\times 10^{-9}$ & $\tabscript{5.22}{+2.02}{-1.70}\times 10^{-9}$ & $\tabscript{9.97}{+4.04}{-3.03}\times 10^{-9}$ \\
$\Upsilon(2S)$    & $\tabscript{2.34}{+0.76}{-1.00}\times 10^{-9}$ & $\tabscript{1.42}{+0.72}{-0.57}\times 10^{-9}$ & $\tabscript{2.62}{+1.39}{-0.91}\times 10^{-9}$ \\
$\Upsilon(3S)$    & $\tabscript{2.13}{+0.76}{-1.13}\times 10^{-9}$ & $\tabscript{0.91}{+0.48}{-0.38}\times 10^{-9}$ & $\tabscript{1.87}{+1.05}{-0.69}\times 10^{-9}$ \\
\hline
\end{tabular}
\end{center}
\end{table}
 
Deviations of the quark Yukawa couplings from SM expectations can lead to significant enhancements in the branching fractions of these radiative decays,
particularly in the bottomonium sector.
Such deviations can arise in BSM theories~\cite{deFlorian:2016spz}.
For instance, the quark masses may not originate entirely from the Higgs mechanism, but could also be
induced by other, subdominant, sources of electroweak symmetry breaking~\cite{Perez:2015aoa}.
Some further examples are the Froggatt--Nielsen mechanism~\cite{Froggatt:1978nt},
the Randall--Sundrum family of models \cite{Randall:1999ee},
the minimal flavour violation framework~\cite{DAmbrosio:2002ex},
the Higgs-dependent Yukawa couplings model~\cite{Giudice:2008uua},
and the possibility of the Higgs boson being a composite pseudo-Goldstone boson~\cite{Dugan:1984hq}.
 
The $Z$ boson production cross section at the LHC~\cite{STDM-2015-03} is approximately $1000$ times larger than the Higgs boson production cross section~\cite{deFlorian:2016spz,Dainese:2703572},
which allows rare $Z$ boson decays to be probed to much smaller branching fractions than Higgs boson decays to the same final state.
Similarly to the Higgs boson decays in Figure~\ref{fig:higgsFeynman}, radiative decays of the $Z$ boson into a vector quarkonium state and a photon receive analogous contributions from direct and indirect amplitudes.
In $Z\rightarrow\mathcal{Q}\,\gamma$ decays, the power corrections in terms of the ratio of the QCD energy scale to the vector-boson mass are small.
As discussed in Ref.~\cite{Grossmann:2015lea}, this allows the light-cone distribution amplitudes (LCDAs) of the mesons
to be probed in a theoretically clean region where power corrections are in control, which is not possible in other applications of the QCD factorisation approach.
These decays have not yet been measured, but recent independent calculations of the SM branching fractions for $Z\to\Jpsi\,\gamma$ and $Z\to\Upsilon(1S,2S,3S)\,\gamma$ are presented in Table~\ref{tab:ZQgammaSM} and are expected to be of order $10^{-8}$ to $10^{-7}$~\cite{Huang:2014cxa,Grossmann:2015lea,Bodwin:2017pzj}.
No value has been calculated for $Z\to\psi(2S)\,\gamma$.
 
\begin{table}[!htbp]
\begin{center}
\caption{Overview of calculations of the Standard Model $Z\rightarrow\mathcal{Q}\,\gamma$ branching fractions.\label{tab:ZQgammaSM}}
\begin{tabular}{|c|c|c|c|}
\hline
Vector            & \multicolumn{3}{c|}{SM branching fraction, $\mathcal{B}\left(Z\rightarrow\mathcal{Q}\,\gamma\right)$} \\ \cline{2-4}
quarkonium state  & Ref.~\cite{Grossmann:2015lea} (2015)            & Ref.~\cite{Huang:2014cxa} (2015)                & Ref.~\cite{Bodwin:2017pzj} (2018) \\ \hline
$\Jpsi$           & $\tabscript{8.02}{+0.46}{-0.44}\times 10^{-8}$ & $\tabscript{9.96}{+1.86}{-1.86}\times 10^{-8}$ & $\tabscript{8.96}{+1.51}{-1.38}\times 10^{-8}$ \\
$\Upsilon(1S)$    & $\tabscript{5.39}{+0.17}{-0.15}\times 10^{-8}$ & $\tabscript{4.93}{+0.51}{-0.51}\times 10^{-8}$ & $\tabscript{4.80}{+0.26}{-0.25}\times 10^{-8}$ \\
$\Upsilon(2S)$    & -                                              & -                                              & $\tabscript{2.44}{+0.14}{-0.13}\times 10^{-8}$ \\
$\Upsilon(3S)$    & -                                              & -                                              & $\tabscript{1.88}{+0.11}{-0.10}\times 10^{-8}$ \\
\hline
\end{tabular}
\end{center}
\end{table}
 
Decays of the Higgs and $Z$ bosons into $\Jpsi$ or $\Upsilon(1S,2S,3S)$ and a photon
were searched for by the ATLAS
Collaboration, initially with
up to $20.3~\ifb$ of data collected at $\sqrt{s}=8~\TeV$~\cite{HIGG-2014-03} and subsequently with up to $36.1~\ifb$ of data collected at $\sqrt{s}=13~\TeV$~\cite{HIGG-2016-23};
the latter search also introduced the study of the $\psi(2S)$ decay channels.
The obtained 95\% confidence level (CL) upper limits on the branching fractions were $3.5\times 10^{-4}$ and $2.0\times 10^{-3}$ for $\HtoJpsiGamma$ and $H\to\psi(2S)\,\gamma$, respectively, and
$\left(4.9,5.9,5.7\right)\times 10^{-4}$ for $H\to\Upsilon(1S,2S,3S)\,\gamma$.
The corresponding 95\% CL upper limits for the analogous $Z$ boson decays were
$2.3\times 10^{-6}, 4.5\times 10^{-6}$ and
$\left(2.8,1.7,4.8\right)\times 10^{-6}$.
The $H\rightarrow J/\psi\,\gamma$ and $Z\rightarrow J/\psi\,\gamma$ decays have also been searched for by the CMS
Collaboration~\cite{CMS-HIG-14-003,CMS-SMP-17-012}, yielding similar upper limits of
$7.6\times 10^{-4}$ and $1.4 \times 10^{-6}$ from $35.9~\ifb$ of data collected at $\sqrt{s}=13~\TeV$.
In addition, the ATLAS Collaboration has searched for the rare Higgs and $Z$ boson decays to light vector mesons
$H(Z)\to\phi\,\gamma$ and $H(Z)\to\rho\,\gamma$~\cite{HIGG-2016-05,HIGG-2016-13}, while the CMS Collaboration has also searched for Higgs boson decays into a $Z$ boson and a $\rho$ or $\phi$ meson~\cite{CMS-HIG-19-012},
and Higgs and $Z$ boson decays into pairs of $J/\psi$ or $\Upsilon(1S,2S,3S)$ mesons~\cite{CMS:2019wch}.
 
This paper describes searches for Higgs and $Z$ boson decays into the
exclusive final states $\Jpsi\,\gamma$, $\psi(2S)\,\gamma$, and $\Upsilon(1S,2S,3S)\,\gamma$, and then
into $\mumu\gamma$, using $139$~fb$^{-1}$ of ATLAS proton--proton ($pp$) collision data collected between 2015 and 2018 at $\rts=13~\TeV$.
Hereafter, where no distinction is relevant, the $\Jpsi$ and $\psi(2S)$ states are collectively denoted by $\psi(nS)$,
and the $\Upsilon(1S,2S,3S)$ states are denoted by $\Upsilon(nS)$.
The results are interpreted in the kappa framework~\cite{Heinemeyer:2013tqa,deFlorian:2016spz} in terms of the ratios $\kappa_c/\kappa_\gamma$ and $\kappa_b/\kappa_\gamma$, where $\kappa_c$, $\kappa_b$ and $\kappa_\gamma$ are the modifiers of the coupling between the Higgs boson and the charm quark, bottom quark, and the effective coupling between the Higgs boson and the photon, respectively.


\section{ATLAS detector and data sample}
\label{sec:atlas}
 
ATLAS~\cite{Aad:2008zzm} is a multipurpose particle
detector with an approximately forward--backward symmetric cylindrical
geometry and near 4$\pi$ coverage in solid angle.\footnote{ATLAS uses a
right-handed coordinate system with its origin at the nominal
interaction point (IP) in the centre of the detector and the
$z$-axis along the beam pipe. The $x$-axis points from the IP to the
centre of the LHC ring, and the $y$-axis points upward. Cylindrical
coordinates $(r,\phi)$ are used in the transverse plane, $\phi$
being the azimuthal angle around the $z$-axis. The pseudorapidity is
defined in terms of the polar angle $\theta$ as
$\eta=-\ln\tan(\theta/2)$.}
It consists of an inner tracking detector, electromagnetic and hadronic calorimeters, and a muon spectrometer.
 
The inner tracking detector (ID) covers the pseudorapidity range
$|\eta|<2.5$ and is surrounded by a thin superconducting solenoid
providing a 2~T magnetic field.
At small radii, a high-granularity silicon pixel detector covers the
vertex region and typically provides four measurements per track.
The first hit is usually in the insertable B-layer, an additional layer installed
in 2015 before $13$~TeV data taking began~\cite{ATLAS-TDR-19,PIX-2018-001-alt}.
It is followed by a silicon microstrip tracker, which provides up to eight
measurement points per track. The silicon detectors are complemented
by a gas-filled straw-tube transition radiation tracker, which enables
radially extended track reconstruction up to $|\eta| = 2.0$ with
typically 35 measurements per track.
 
Electromagnetic (EM) calorimetry within $|\eta|<3.2$ is
provided by barrel and endcap high-granularity lead/liquid-argon (LAr)
EM calorimeters with an additional thin LAr presampler covering
$|\eta| < 1.8$ to correct for energy loss in upstream material;
for $|\eta| < 2.5$ the EM calorimeter is divided into three layers in depth.
A steel/scintillator-tile calorimeter provides hadronic calorimetry for
$|\eta|<1.7$. LAr technology with copper as absorber is
used for the hadronic calorimeters in the endcap region,
$1.5<|\eta|<3.2$.  The solid-angle coverage is completed with forward
copper/LAr and tungsten/LAr calorimeter modules in $3.1 < |\eta| <
4.9$ optimised for EM and hadronic measurements, respectively.
 
The muon spectrometer (MS) surrounds the calorimeters and features
trigger and high-precision tracking chambers measuring the
deflection of muons in a magnetic field provided by three air-core
superconducting toroidal magnets.
The field integral of the toroids ranges between \num{2.0} and \SI{6.0}{\tesla\metre}
across most of the detector.
The precision chamber system covers the
region $|\eta| < 2.7$ with three layers of monitored drift tubes,
complemented by cathode strip chambers in the forward regions. The muon
trigger system covers the range $|\eta| < 2.4$, featuring resistive plate
chambers in the barrel and thin gap chambers in the endcap regions.
 
A two-level trigger and data acquisition system is used to record
events for offline analysis~\cite{TRIG-2016-01}. The first-level trigger
is implemented in hardware and uses a subset of detector information.
This is followed by a software-based high-level trigger which outputs events for permanent storage at an average rate of 1~kHz, reduced from the maximum first-level rate of 100 kHz.
An extensive software suite~\cite{ATL-SOFT-PUB-2021-001} is used in the reconstruction and analysis of real and simulated data,
in detector operations, and in the trigger and data acquisition systems of the experiment.
 
Data samples considered in this analysis were
collected during stable beam conditions with all relevant detector systems functional~\cite{DAPR-2018-01} and were recorded by a combination
of triggers requiring a photon and either one or two muons in the event.
Due to the increasing instantaneous luminosity, the transverse momentum thresholds and identification requirements were modified during the
data-taking periods.
Available throughout the data collection period was a trigger requiring a photon fulfilling the `medium' identification criteria~\cite{Aad:2019tso} and transverse
momentum $\ptgamma$ greater than $25~\GeV$, and at least one muon identified at the first-level trigger with
$\ptmu$ greater than $24~\GeV$.
In the 2015 and 2016 data collection periods, this was complemented with a trigger requiring a photon fulfilling the `loose' identification criteria~\cite{Aad:2019tso} and
$\ptgamma>35~\GeV$, and at least one muon identified at the high-level trigger with
$\ptmu>18~\GeV$.
During the 2017 and 2018 data collection, a trigger requiring an isolated photon fulfilling the `tight' identification criteria~\cite{Aad:2019tso} and $\ptgamma>35~\GeV$ and
at least one muon identified at the high-level trigger with
$\ptmu>18~\GeV$, and a trigger requiring a `loose'  photon with $\ptgamma>35~\GeV$, a muon identified at the first-level trigger with $\ptmu>15~\GeV$ and
an additional muon identified at the high-level trigger with
$\ptmu>2~\GeV$ were used. For candidate signal events that fulfil the analysis selection,
described in Section~\ref{sec:selection}, the trigger efficiency exceeds 97\% in all cases.
The uncertainty in the trigger efficiency is estimated to be 0.8\%~\cite{TRIG-2018-05,TRIG-2018-01}.
After applying trigger and data-quality requirements, the integrated luminosity of the data sample used in this search
corresponds to $139.0\pm 2.4~\ifb$, obtained using the LUCID-2 detector~\cite{LUCID2} for the primary luminosity measurements and
with the uncertainty in the integrated luminosity derived using the method
described in Refs.~\cite{DAPR-2013-01,ATLAS-CONF-2019-021}.
The average number of $pp$ interactions per bunch crossing ranged
from about 13 in 2015 to about 39 in 2018.


\section{Event selection}
\label{sec:selection}
 
In addition to the trigger and data-quality requirements, selection criteria based on geometric acceptance, event kinematics, isolation, and vertex quality are imposed to select
candidate $H(Z)\rightarrow\cal{Q}\,\gamma\rightarrow\mumu\,\gamma$ events.
Reconstructed muon candidates must be either `segment-tagged', where an ID track matches at least one track segment in the MS,
or `combined', where an ID track matches a full track in the MS~\cite{PERF-2014-05}.
Muons must also have an absolute value of pseudorapidity $|\etamu|<2.5$ and a transverse momentum $\ptmu>3~\GeV$.
Pairs of oppositely charged muons are fitted to a common vertex with the track parameter uncertainties taken into account~\cite{Kostyukhin:2003daa}.
Pairs satisfying a loose $\chi^2$ requirement for this fit are combined to reconstruct candidate $\cal{Q}\to\mumu$ decays.
The higher-$\pt$ muon in each pair, called the leading muon, is required to have $\ptmu>18~\GeV$.
Dimuons with a mass $m_\mumu$ between $2.4~\GeV$ and $4.3~\GeV$ are selected as $\psi(nS)\to\mumu$ candidates,
and those with a mass $8.0~\GeV<m_\mumu<12.0~\GeV$ are selected as $\Upsilon(nS)\to\mumu$ candidates.
To improve the sensitivity of the $\Upsilon(nS)\gamma$ analysis in resolving the individual $\Upsilon(nS)$ states,
events are classified into two exclusive categories based upon muon pseudorapidity to account for differences in resolution
across the ATLAS detector.
Events where both muons satisfy $|\etamu|<1.05$ are placed in the higher-resolution barrel (B) category,
otherwise they are placed in the endcap (EC) category.
 
Candidate $\cal{Q}\to\mumu$ decays are subjected to further isolation and vertex-quality requirements.
The sum of the $\pt$ of ID tracks within a cone of variable size $\Delta R = \sqrt{(\Delta\phi)^{2} + (\Delta\eta)^{2}} = \min\{10~\GeV/(\ptmu~[\GeV]), 0.3\}$
around the leading muon is required to be less than 6\% of the $\mathcal{Q}$ candidate's transverse momentum, $\pt^{\mumu}$~\cite{Aad:2020gmm}.
This sum excludes the ID track associated with the leading muon itself,
as well as the ID track associated with the subleading muon if it falls inside the $\Delta R$ cone.
To mitigate the effects of multiple $pp$ interactions in the same or
neighbouring bunch crossings, only ID tracks that originate from the primary vertex are considered,
which is defined as the reconstructed vertex with the highest $\Sigma{\,\pt}^2$
of all associated tracks used in the formation of the vertex.
To reject contributions from events involving  $b$-hadron decays which result in displaced vertices,
the vector leading from the primary vertex to the dimuon vertex is projected
onto the direction of the $\cal{Q}$ candidate's transverse momentum, and the signed projection $L_{xy}$
is required to be smaller than three times its uncertainty, $\sigma_{L_{xy}}$, such that $|L_{xy}/\sigma_{L_{xy}}| < 3$.
 
Photons are reconstructed from clusters of energy in the electromagnetic calorimeter.
Clusters that match ID tracks consistent with the hypothesis of a photon conversion into $\ee$
are classified as converted photon candidates, whilst clusters that have no matching ID tracks
are classified as unconverted candidates~\cite{Aad:2019tso}.
Reconstructed photon candidates are required to satisfy the `tight' photon identification criteria~\cite{Aad:2019tso},
have absolute pseudorapidity $|\etagamma|<2.37$, excluding the calorimeter barrel/endcap transition region $1.37<|\etagamma|<1.52$,
and have transverse momentum $\ptgamma>35~\GeV$.
Further track- and calorimeter-isolation requirements are imposed to suppress contamination from jets:
the sum of the \pt of all ID tracks originating from the primary vertex and within $\Delta R = 0.2$ of the photon direction,
excluding any associated with the reconstructed photon, is required to be less than $5\%$ of $\ptgamma$;
the sum of the \pt of the calorimeter energy clusters within $\Delta R = 0.4$ of the photon direction,
excluding the energy cluster of the reconstructed photon, is required to be less than $\left(2.45~\GeV + 0.022 \times \ptgamma~[\GeV]\right)$.
The calorimeter isolation variable is corrected to account for contributions associated with other $pp$ interactions in the same bunch crossing~\cite{Aad:2019tso}.
 
Combinations of a $\cal{Q}\to\mumu$ candidate and a photon that satisfy $\Delta\phi({\cal{Q}},\gamma)>\pi/2$ are retained for further analysis: this requirement suppresses contributions from events where the quarkonium and photon candidates have a small angular separation.
If multiple combinations are possible, a situation that arises in fewer than $2\%$ of events that pass the trigger requirements,
the combination of the highest-\pt photon and the $\cal{Q}$ candidate with an invariant mass closest to that of the \Jpsi meson is retained.
To maintain a common event selection for the Higgs and $Z$ boson analyses, while ensuring near-optimal sensitivity for both,
a variable $\pt^{\mumu}$ threshold that depends on the invariant mass of the $\cal{Q}\gamma$ system, $m_{\mumu\gamma}$, is applied.
For the $\psi(nS)\to\mumu$ ($\Upsilon(nS)\to\mumu$) candidates, the $\pt^{\mumu}$ threshold is $40~\GeV$ ($34~\GeV$) for $m_{\mumu\gamma}\leq91~\GeV$,
and $54.4~\GeV$ ($52.7~\GeV$) for $m_{\mumu\gamma}\geq140~\GeV$.
In the region $91~\GeV<m_{\mumu\gamma}<140~\GeV$, the $\pt^{\mumu}$ threshold varies linearly between its fixed values outside this region.
The $\pt^{\mumu}$ thresholds are chosen to optimise the significance of potential signals at the Higgs and $Z$ boson masses.


\section{Signal modelling}
\label{sec:signalmodel}
 
The expected $H(Z)\rightarrow \cal{Q}\gamma$ signals were modelled using Monte Carlo (MC) simulated events to extract analytical approximations of the final observables.
Several Higgs boson production modes were considered in the simulation of the $H\rightarrow \cal{Q}\gamma$ events.
In order of decreasing production cross section, these are gluon--gluon fusion ($ggH$), vector-boson fusion (VBF), and associated production of a $H$ boson with a $Z$ boson ($ZH$), a $W^{\pm}$ boson ($WH$) or a $t\bar{t}$ top-quark pair ($t\bar{t}H$).
Explicit simulation of the $t\bar{t}H$ Higgs boson production mechanism is an addition with respect to the analysis strategy of the $36.1~\ifb$ study~\cite{HIGG-2016-23}.
The contribution of Higgs bosons produced in association with a $b\bar{b}$ bottom-quark pair ($b\bar{b}H$) was not directly simulated, but was included by scaling up the production cross section used to normalise the $ggH$ sample.
This assumes that the efficiency for $b\bar{b}H$ events is equal to that for $ggH$; their inclusion increases the signal yield by less than 1\%.
For the $Z\rightarrow \cal{Q}\gamma$ signal events, $Z$ boson production was modelled inclusively.
 
The \progname{Powheg\,Box}\,v2 MC event generator~\cite{powheg1,powheg2,powheg3,powheg4,powheg5}
was used to model the $ggH$ and VBF Higgs boson production mechanisms,
calculated up to next-to-leading order (NLO) in $\alpha_\text{s}$.
\progname{Powheg\,Box} was interfaced with \progname{Pythia\,8.212}~\cite{Sjostrand:2007gs,Sjostrand:2006za},
which used the CTEQ6L1 parton distribution functions~\cite{Pumplin:2002vw} and
a set of tuned parameters called the AZNLO tune~\cite{STDM-2012-23} to model the parton shower, hadronisation, and underlying event.
The \progname{Pythia\,8.212} event generator was used to model the $ZH$ and $WH$ production mechanisms,
with \progname{NNPDF2.3lo} parton distribution functions~\cite{Ball:2012cx} and the A14 tune~\cite{ATL-PHYS-PUB-2014-021} for hadronisation and the underlying event.
The \progname{MadGraph5\_aMC@NLO}\,2.2.2~\cite{Alwall:2014hca} event generator, interfaced with \progname{Pythia\,8.212} for the parton shower,
was used to model the $t\bar{t}H$ production mechanism, using the same parton distribution functions and event tune as the $ZH$ and $WH$ samples.
Inclusive $Z$ boson production was modelled with the same event generators, parton distribution functions, and event tune as the $ggH$ and VBF Higgs boson production mechanisms.
The subsequent decays of the Higgs and $Z$ bosons into $J/\psi\,\gamma$, $\psi(2S)\,\gamma$ and $\Upsilon(1S,2S,3S)\,\gamma$ were simulated as a cascade of two-body
decays, where the explicit simulation of $\psi(2S)\,\gamma$ decays is an addition to the analysis strategy compared to the $36.1~\ifb$ study~\cite{HIGG-2016-23}, which modelled these events by using $J/\psi\,\gamma$ samples.
Interference effects between $J/\psi\,\gamma$ produced in $Z$ boson decays and non-resonant QCD gluon-mediated processes~\cite{Jones:2020bvu} are expected to be small and were neglected.
The generated events were passed through the detailed $\GEANT$ simulation of the ATLAS detector~\cite{GEANT4,SOFT-2010-01}
and processed with the same software as used to reconstruct the data.
Conditions in the ATLAS detector, such as the average number of $pp$ interactions per bunch crossing, changed throughout Run~2.
Separate samples were produced with 2015--2016 conditions, 2017 conditions, and 2018 conditions,
where each sample is normalised according to the integrated luminosity of the corresponding run period.
The effect of multiple interactions in the same or neighbouring bunch crossings (pile-up) was modelled by overlaying each simulated hard-scattering event with inelastic $pp$ events generated with \progname{Pythia\,8.186} using the \progname{NNPDF2.3lo} parton distribution functions and the A3 tune~\cite{ATLAS:2016puo}.

The production rates
for the SM Higgs boson with $\mH = 125~\GeV$, obtained from the CERN Yellow Reports~\cite{deFlorian:2016spz,Dainese:2703572}, are assumed
throughout this analysis.
The $ggH$ sample is normalised such that it reproduces the total cross section predicted by a next-to-next-to-next-to-leading-order (N$^3$LO) QCD calculation with NLO electroweak corrections applied~\cite{Anastasiou:2015ema,Anastasiou:2016cez,Actis:2008ug,Anastasiou:2008tj}.
The VBF sample is normalised to an approximate next-to-next-to-leading-order (NNLO) QCD cross section with NLO electroweak corrections applied~\cite{Ciccolini:2007jr,Ciccolini:2007ec,Bolzoni:2010xr}.
The samples for the associated production of a $W$ or $Z$ boson with a Higgs boson are normalised to cross sections calculated at NNLO in QCD with NLO electroweak corrections~\cite{Brein:2003wg,Denner:2011id} including the NLO QCD corrections~\cite{Altenkamp:2012sx} for $gg\to ZH$.
The production of $t\bar{t}H$ is normalised to cross sections calculated at NLO in QCD with NLO electroweak corrections~\cite{deFlorian:2016spz}.
The production cross section used to scale the $ggH$ sample to account for the $b\bar{b}H$ mechanism was calculated at a mix of NNLO and NLO accuracy in QCD, with no electroweak corrections~\cite{deFlorian:2016spz}.
The production rate for the $Z$ boson is normalised to the total cross section obtained from a measurement by the ATLAS Collaboration using $81~\ipb$ of $\sqrt{s}=13$~\TeV\ data~\cite{STDM-2015-03}.
The branching fractions for the decays $\cal{Q}\to\mumu$ used in signal normalisation are taken from the Review of Particle Physics~\cite{PDG},
as are the inclusive branching fractions for $t$-quark decays to hadrons or leptons in the normalisation of the $t\bar{t}H$ samples.
The effects of meson polarisation on the dimuon kinematic distributions are accounted for via a reweighting of the simulated events, which are initially simulated without polarisation.
The quarkonium states in the decay of the Higgs boson are expected to be transversely polarised;
the quarkonium states in the decay of the $Z$ boson are expected to be longitudinally polarised, due to a vanishing contribution from the transversely polarised meson~\cite{Huang:2014cxa}.
Accounting for meson polarisation results in a $2\%$--$3\%$ decrease in Higgs boson signal efficiency and a $9\%$--$10\%$ increase in $Z$ boson signal efficiency.
 
Figure~\ref{fig:pt_ph_muons} shows the generator-level photon and muon \pt distributions from the simulated $H(Z)\rightarrow\cal{Q}\gamma$ signal events.
For the $\Jpsi\,\gamma\to\mumu\,\gamma$ and $\psi(2S)\,\gamma\to\mumu\,\gamma$ final states, the total signal efficiency is
19\% for the Higgs boson decays and 10\% for the $Z$~boson decays.
These values take into account the trigger, reconstruction, identification, and isolation efficiencies, as well as the kinematic acceptance.
The corresponding values for the $\Upsilon(1S,2S,3S)\,\gamma\to\mumu\,\gamma$ final states are 21\% and 13\%.
The difference in efficiency between the Higgs and $Z$ boson decays arises
primarily from the softer photon and muon \pt distributions associated with
$Z\to\cal{Q}\,\gamma$ production, as seen by comparing Figures~\ref{fig:pt_ph_muons_Hjpsi} and \ref{fig:pt_ph_muons_Zjpsi} for the $\Jpsi\,\gamma$ case,
Figures~\ref{fig:pt_ph_muons_Hpsi2S} and \ref{fig:pt_ph_muons_Zpsi2S} for the $\psi(2S)\,\gamma$ case,
and Figures~\ref{fig:pt_ph_muons_Hupsi} and \ref{fig:pt_ph_muons_Zupsi} for the $\Upsilon(nS)\,\gamma$ case.
 
\begin{figure}[!h]
\centering
\subfigure[$H\to\Jpsi\,\gamma$ \label{fig:pt_ph_muons_Hjpsi}]{\includegraphics[width=0.3\textwidth]{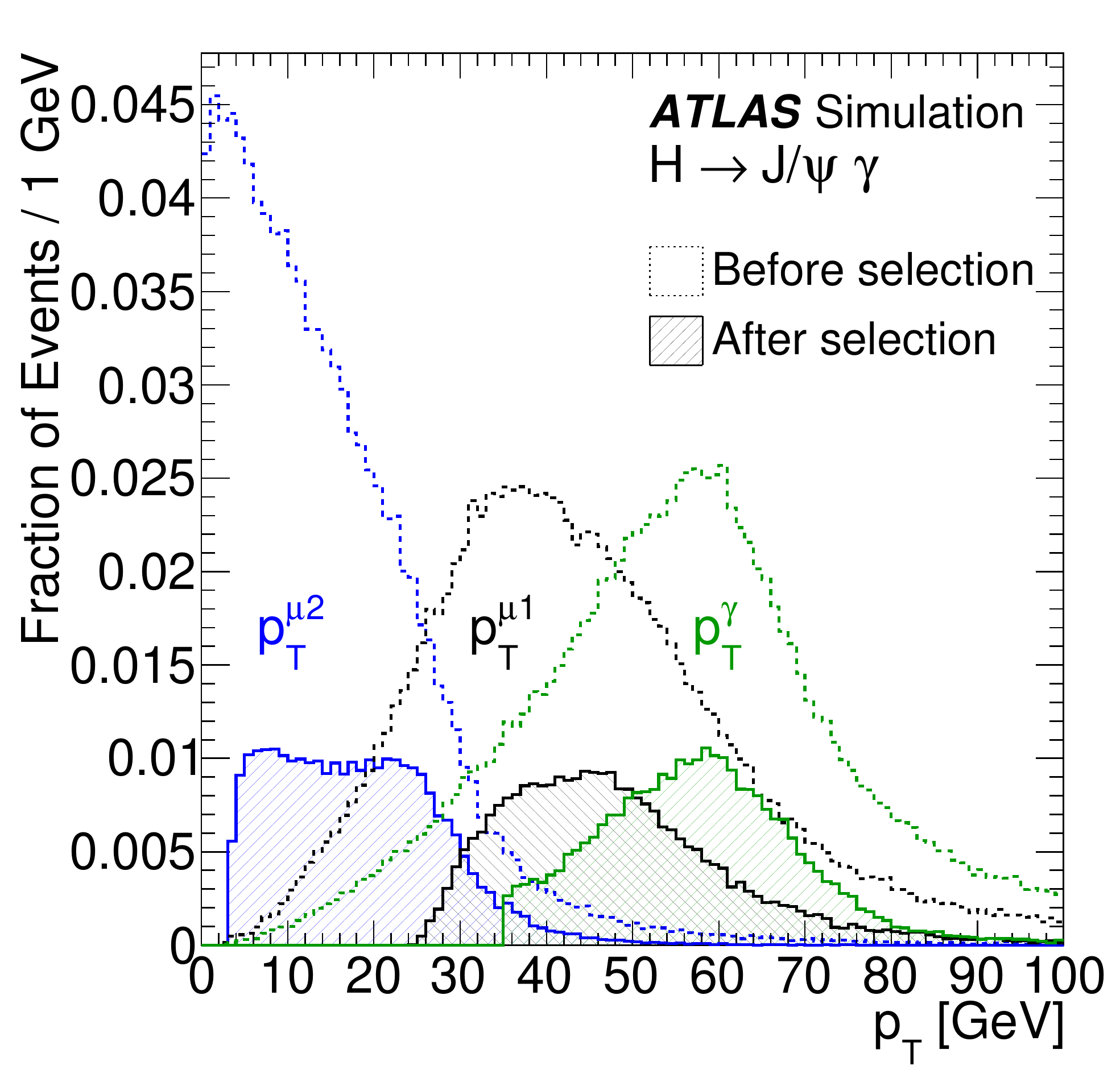}}
\subfigure[$H\to\psi(2S)\,\gamma$ \label{fig:pt_ph_muons_Hpsi2S}]{\includegraphics[width=0.3\textwidth]{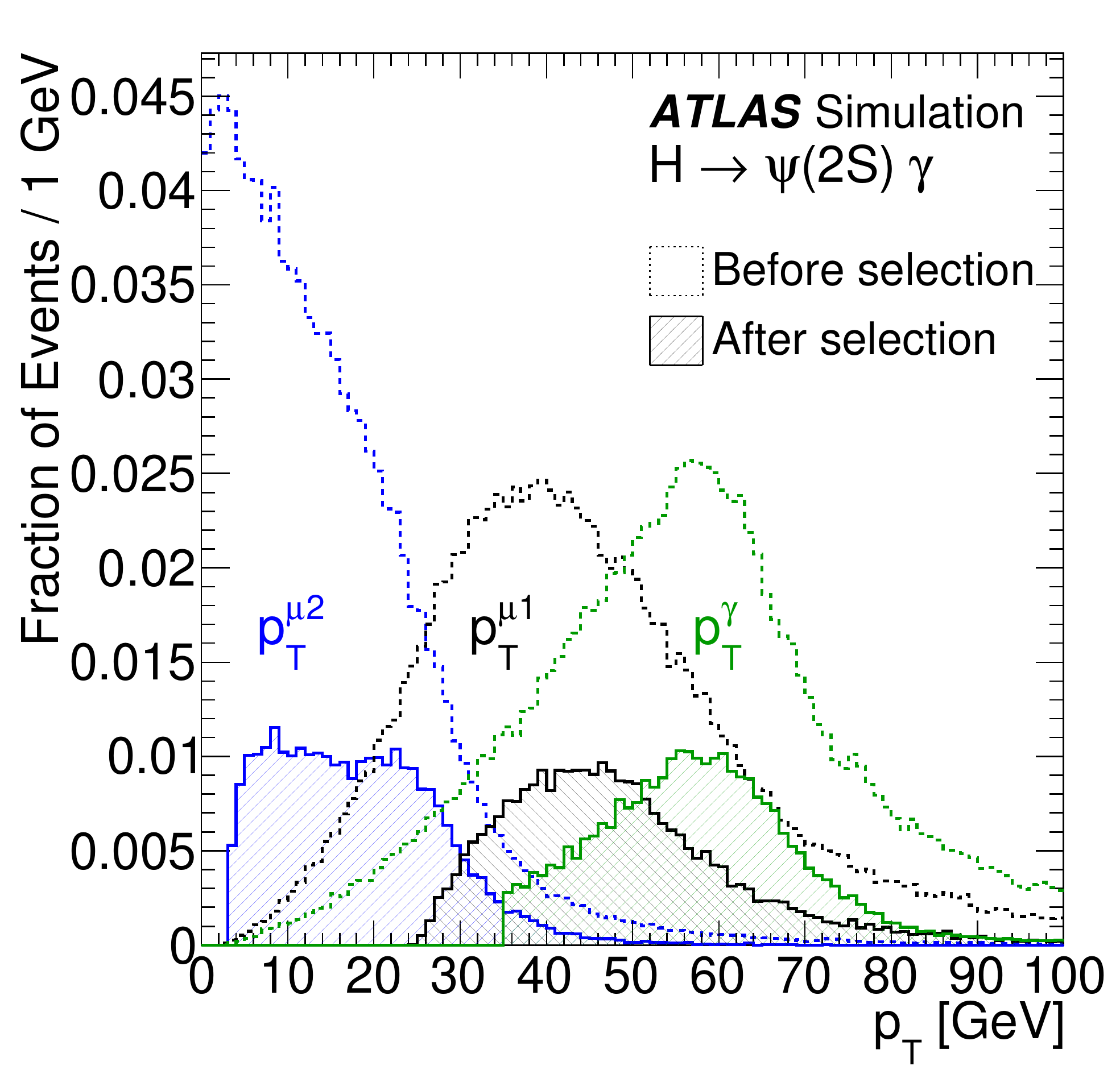}}
\subfigure[$H\to\Upsilon(nS)\,\gamma$ \label{fig:pt_ph_muons_Hupsi}]{\includegraphics[width=0.3\textwidth]{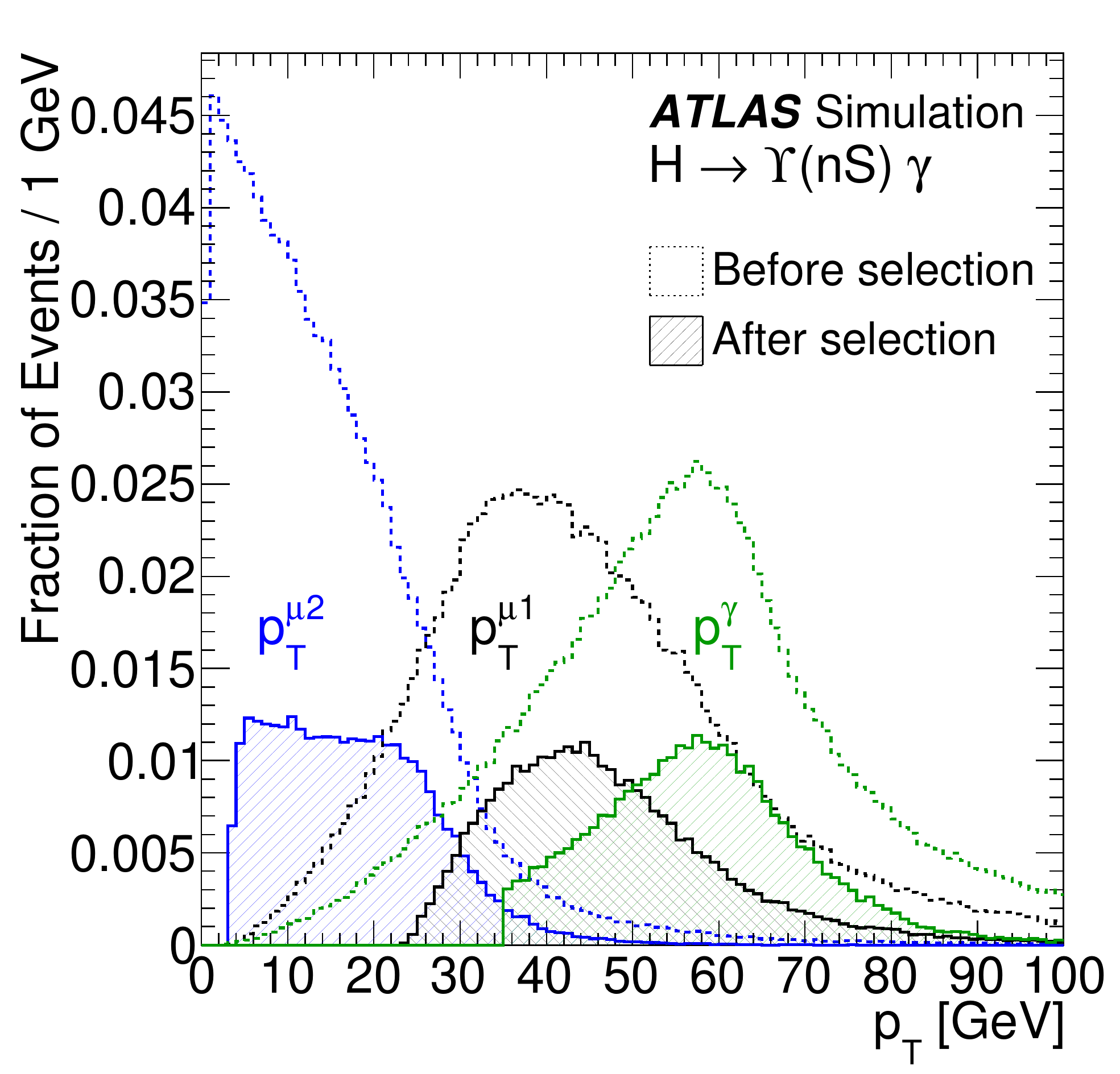}}
 
\subfigure[$Z\to\Jpsi\,\gamma$ \label{fig:pt_ph_muons_Zjpsi}]{\includegraphics[width=0.3\textwidth]{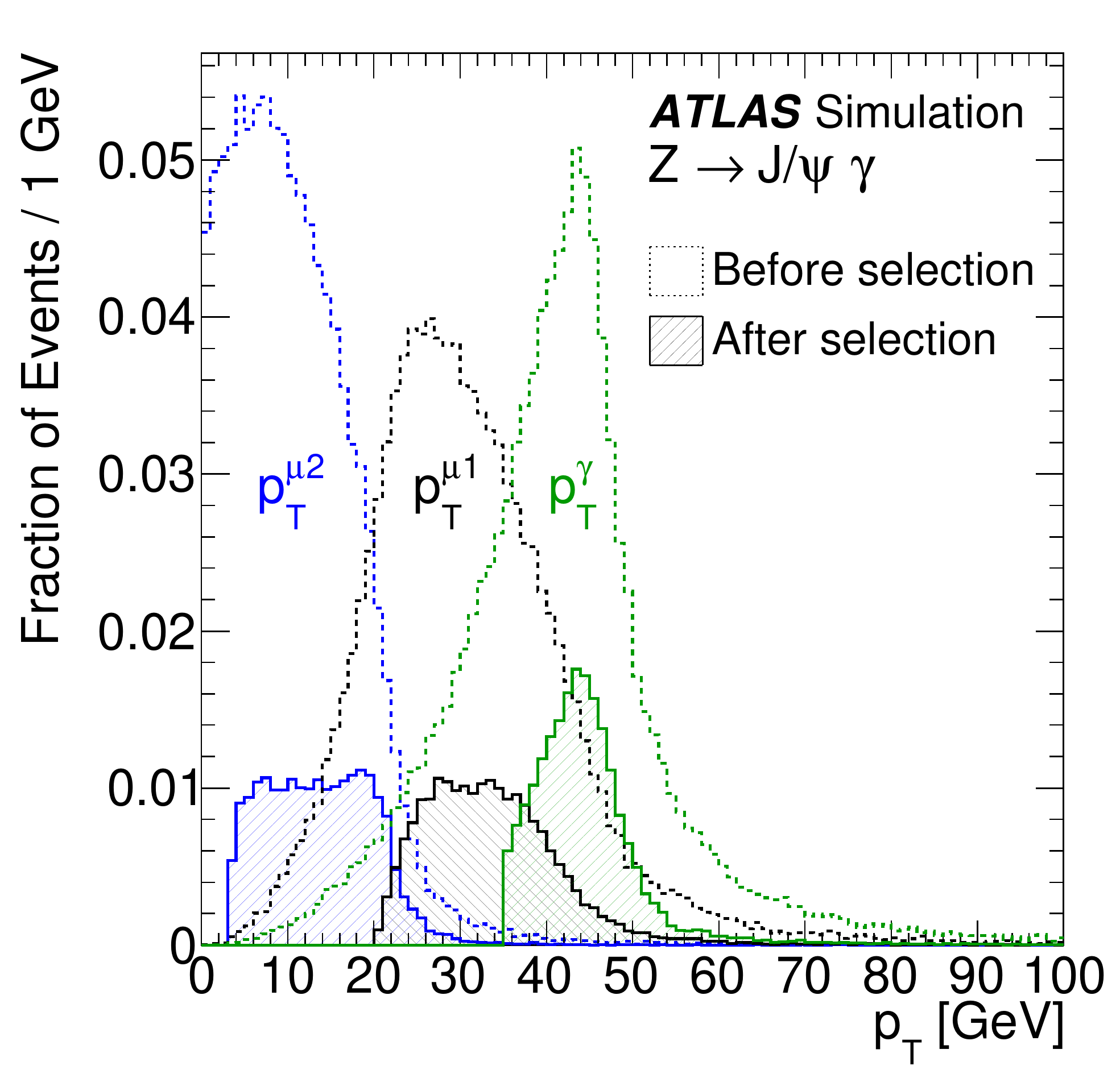}}
\subfigure[$Z\to\psi(2S)\,\gamma$ \label{fig:pt_ph_muons_Zpsi2S}]{\includegraphics[width=0.3\textwidth]{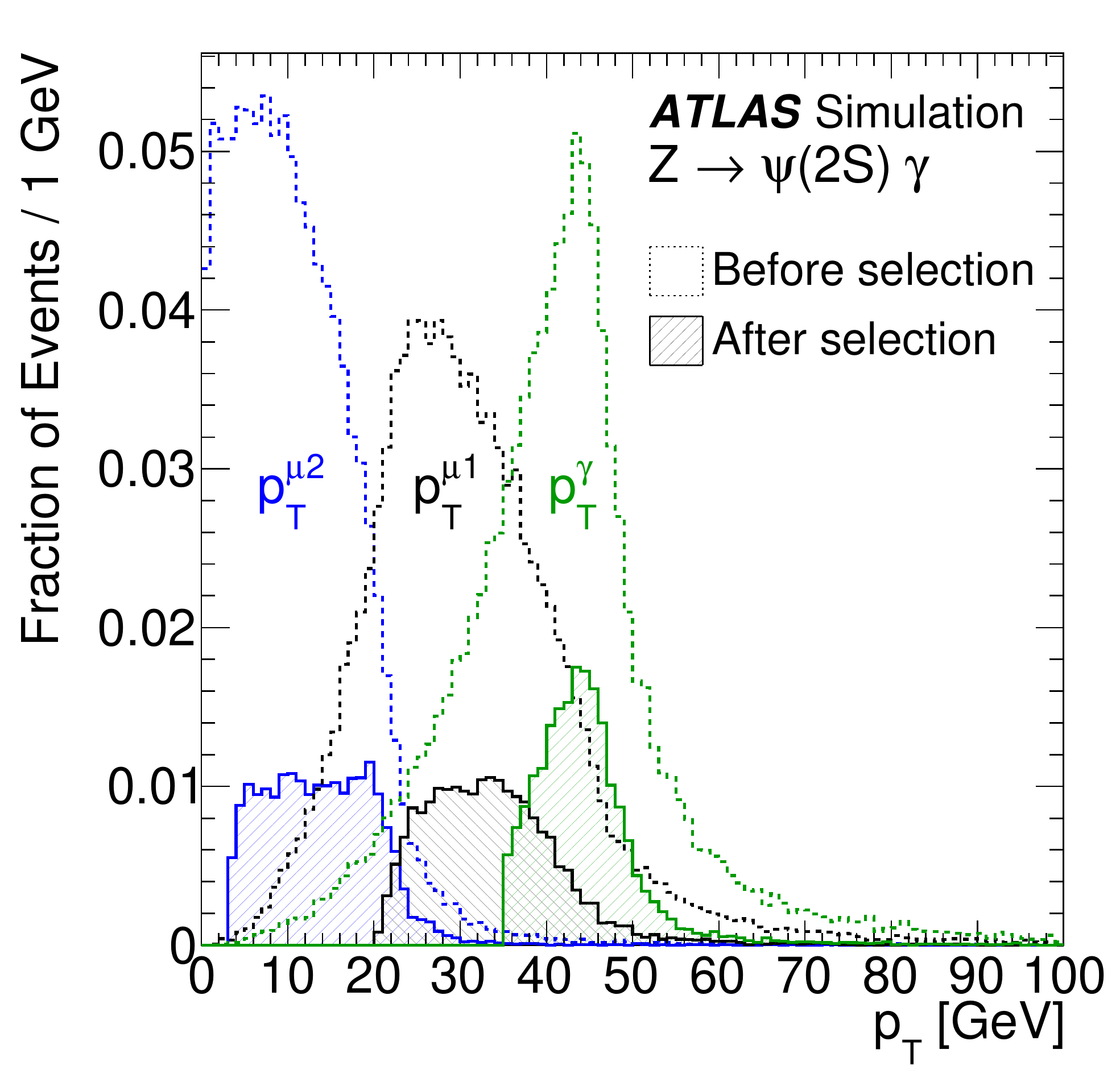}}
\subfigure[$Z\to\Upsilon(nS)\,\gamma$ \label{fig:pt_ph_muons_Zupsi}]{\includegraphics[width=0.3\textwidth]{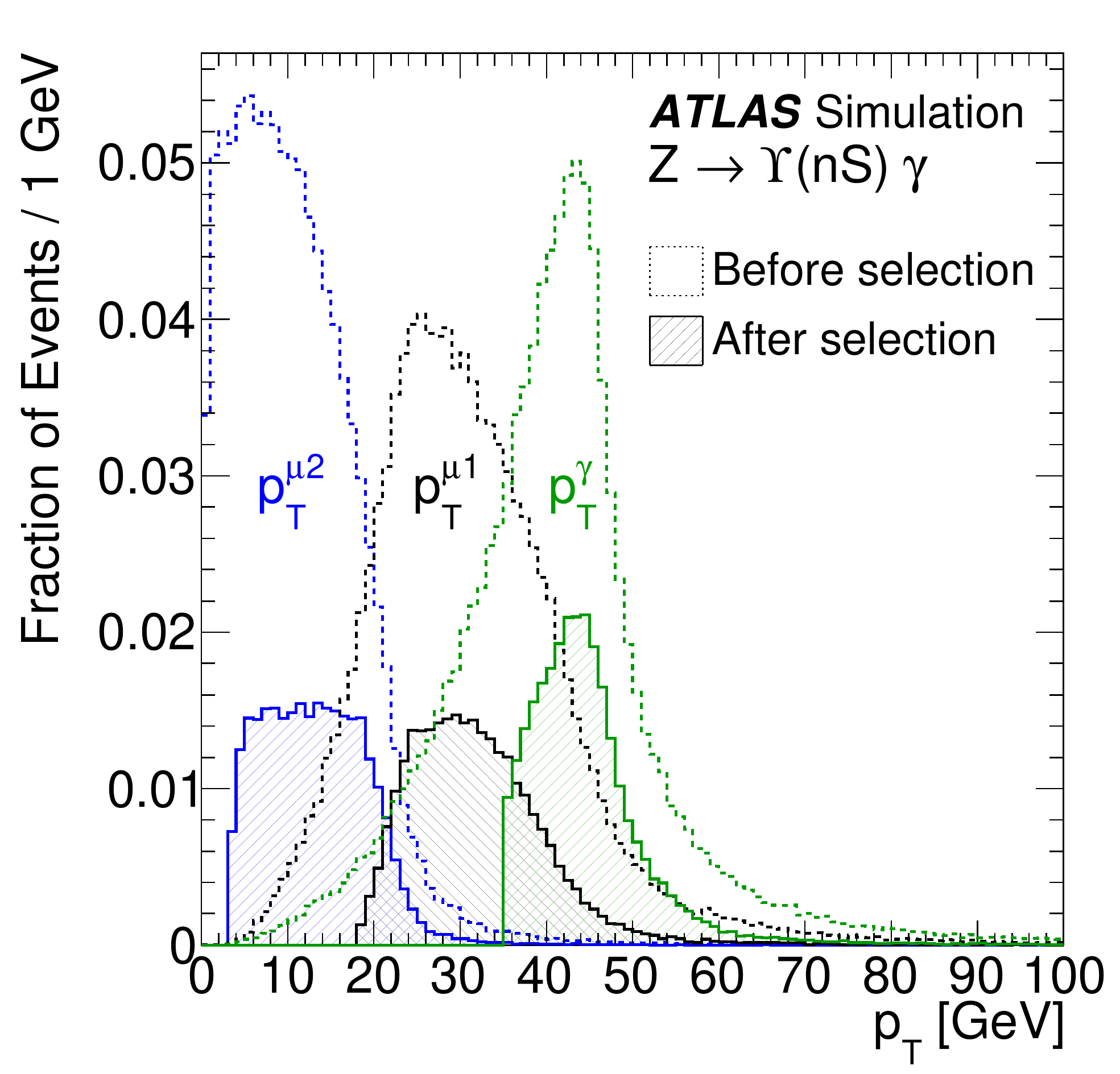}}
\caption{Generator-level transverse momentum ($\pt$) distributions of the photon and muons for \subref{fig:pt_ph_muons_Hjpsi} $H\to\Jpsi\,\gamma$, \subref{fig:pt_ph_muons_Hpsi2S} $H\to\psi(2S)\,\gamma$, \subref{fig:pt_ph_muons_Hupsi} $H\to\Upsilon(nS)\,\gamma$, \subref{fig:pt_ph_muons_Zjpsi} $Z\to\Jpsi\,\gamma$, \subref{fig:pt_ph_muons_Zpsi2S} $Z\to\psi(2S)\,\gamma$ and \subref{fig:pt_ph_muons_Zupsi} $Z\to\Upsilon(nS)\,\gamma$ simulated events, respectively.
The dashed-line distributions with a clear fill show the events at generator level which fall within the analysis geometric acceptance (both muons are required to have $|\etamu|<2.5$, while the photon is required to have $|\etagamma|<2.37$, excluding the region $1.37<|\etagamma|<1.52$), and are each normalised to unity.
The solid-line distributions with a hatched fill show the fraction of these events which pass the full analysis event selection described in Section~\ref{sec:selection}.
The relative difference between the two sets of distributions corresponds to the effects of reconstruction, trigger, and event selection efficiencies.
The leading muon candidate is denoted by $\pt^{\mu 1}$ (black), the subleading candidate by $\pt^{\mu 2}$ (blue), and the photon candidate is denoted by $\pt^{\gamma}$ (green).\label{fig:pt_ph_muons}}
\end{figure}
 
The $m_{\mumu\,\gamma}$ resolution is 1.6\%--1.8\% for both the Higgs and $Z$~boson decays.
For each of the final states, a two-dimensional probability density function (PDF) is used to model the signal in $m_{\mumu\gamma}$ and $m_{\mumu}$.
The Higgs boson signals are modelled with the sum of two bivariate Gaussian distributions, which describe the approximately $60\%$ correlations between the two mass variables in each decay channel as well as the effects of detector resolution.
For $Z$ boson decays, since the $Z$ boson natural width is comparable to the detector resolution, the correlation between $m_{\mumu\gamma}$ and $m_{\mumu}$ is small, of order $10\%$, and is neglected, and the two mass distributions are treated as uncorrelated. The $m_{\mumu\gamma}$ distributions of the $Z$ boson signals are modelled with the sum of two Voigtian PDFs corrected with a mass-dependent efficiency factor, which accounts for the changing selection efficiency along the $Z$ lineshape that arises from the kinematic requirements described in Section~\ref{sec:selection}.
The Voigtian shape is a convolution of a Breit--Wigner distribution, which describes the natural width of the $Z$ boson, and a Gaussian distribution, which describes detector resolution effects.
The $m_{\mumu}$ distributions of the $Z$ boson decays are modelled with a sum of two Gaussian PDFs, where the peak value and resolution parameters of the PDFs are fixed to the values obtained in a fit to the simulated event samples.
 
Systematic uncertainties in the signal yield and inferred branching fraction of the $H$ and $Z$ boson decays are considered.
Uncertainties in the Higgs boson production cross sections total 5.8\%~\cite{deFlorian:2016spz,Dainese:2703572}.
The uncertainty in the measured $Z$ boson production cross section is 2.9\%~\cite{STDM-2015-03}, where the luminosity component of the uncertainty in the $Z$ boson production cross-section measurement is treated as completely uncorrelated with the integrated luminosity uncertainty of data set used in this search.
The uncertainty in the integrated luminosity is estimated to be $1.7\%$~\cite{DAPR-2013-01,ATLAS-CONF-2019-021}, using primary luminosity measurements from the LUCID-2 detector~\cite{LUCID2}.
The uncertainty in the acceptance of the Higgs boson signal
due to the choice of MC generator parameter values, parton distribution functions, set of tuned parameters for the underlying event, and parton showering, is estimated by
studying how different choices affect the acceptance at generator level.
The total uncertainty in the Higgs boson signal acceptance is estimated to be 1.8\%.
For the $Z$ boson, the respective signal acceptance uncertainty is determined to be 1.0\% by comparing the $Z$ boson kinematic distributions in simulated events with measurements in data~\cite{STDM-2018-14}.
Trigger efficiencies for photons are determined from samples enriched with
$Z\to\ee$ events in data~\cite{ATLAS-CONF-2011-114}.
The photon trigger efficiency is estimated to contribute a systematic uncertainty of 0.8\%
to the expected signal yields~\cite{TRIG-2018-05,TRIG-2018-01}.
Photon identification efficiencies are determined using the enriched $Z\to\ee$ event samples,
as well as inclusive photon events and $Z\to\ell^+\ell^-\gamma$ events~\cite{Aaboud:2016yuq,ATL-PHYS-PUB-2016-014}.
The photon identification
efficiency uncertainties, for both the converted and unconverted
photons, are 1.7\%--1.9\% for the Higgs and
$Z$~boson signals. The effect of the muon reconstruction and identification efficiency uncertainty
is 2.2\%--2.4\%~\cite{PERF-2015-10}.
The photon energy scale uncertainty, determined from $Z\to\ee$ events
and validated using $Z\to\ell^+\ell^- \gamma$
events~\cite{ref:run1-egamma-calib,ATL-PHYS-PUB-2016-015}, is propagated through the
simulated samples as a function of $\etagamma$ and $\ptgamma$.
The uncertainty associated with the photon energy
scale and resolution in the simulation has a 0.1\%--0.2\% effect on the Higgs and $Z$ boson signal yields. Similarly, the systematic uncertainty associated
with the scale of the muon momentum measurement has a 0.1\%--0.5\% effect on the signal yields~\cite{PERF-2015-10}.
To assess any effect on the expected signal yield from imperfect modelling of pile-up, the average number of pile-up interactions is varied in the simulation; the corresponding uncertainty is 0.7\%--1.1\%.
These systematic uncertainties in the expected signal yields are summarised in Table~\ref{tab:systematics}.
The effect of the energy and momentum scale and resolution uncertainties on the Higgs and $Z$ boson signal shapes is negligible.
 
\begin{table}[h]
\centering
\caption{Summary of the systematic uncertainties in the expected signal yields. In the case of the $H$ decays, the `signal acceptance' uncertainty was estimated through parton distribution function variations, scale variations and variations of the underlying-event tune. For the $Z$ decays, the `signal acceptance' uncertainty was estimated by comparing simulation to data.\label{tab:systematics}}
\begin{tabular}{l|cccc}
\hline
\multirow{2}{*}{Source of systematic uncertainty} & \multicolumn{4}{c}{Signal yield uncertainty} \\
\cline{2-5}
& $H\to\psi(nS)\,\gamma$ & $H\to\Upsilon(nS)\,\gamma$ & $Z\to\psi(nS)\,\gamma$ & $Z\to\Upsilon(nS)\,\gamma$ \\ \hline
Total cross section & 5.8\%~~ & 5.8\%~~ & 2.9\%~~ & 2.9\%~~ \\
Integrated luminosity & 1.7\%~~ & 1.7\%~~ & 1.7\%~~ & 1.7\%~~ \\
Signal acceptance & 1.8\%~~ & 1.8\%~~ & 1.0\%~~ & 1.0\%~~ \\
Muon reconstruction & 2.3\%~~ & 2.2\%~~ & 2.4\%~~ & 2.4\%~~ \\
Photon identification & 1.7\%~~ & 1.7\%~~ & 1.9\%~~ & 1.9\%~~ \\
Pile-up uncertainty & 0.8\%~~ & 0.7\%~~ & 1.1\%~~ & 1.1\%~~ \\
Trigger efficiency & 0.7\%~~ & 0.7\%~~ & 0.8\%~~ & 0.8\%~~ \\
Photon energy scale & 0.1\%~~ & 0.1\%~~ & 0.2\%~~ & 0.2\%~~ \\
Muon momentum scale & 0.1\%~~ & 0.1\%~~ & 0.5\%~~ & 0.2\%~~ \\
Muon momentum resolution (ID) & $<$0.01\%~~~ & 0.01\% & 0.06\% & 0.02\% \\
Muon momentum resolution (MS) & 0.02\% & 0.01\% & 0.04\% & 0.01\% \\ \hline
\end{tabular}
\end{table}


\section{Background modelling}
\label{sec:backgroundmodel}
 
The background is considered to be composed of two distinct contributions which are modelled separately in this analysis.
The first is an exclusive contribution originating from $\mumu\gamma$ events produced via the Drell--Yan process, where a highly energetic photon typically arises from final-state radiation.
The second, which is the dominant background, is an inclusive contribution mostly from multi-jet and $\gamma$+jet events involving dimuon or $\mathcal{Q}$ production.
The exclusive background is modelled with simulation, similarly to the signal model in Section~\ref{sec:signalmodel},
whilst the inclusive background is modelled with a data-driven technique, discussed in detail in Ref.~\cite{Chisholm:2021pdn}.
The background is modelled independently for the $\psi(nS)$ dimuon mass region, and the B and EC categories of the $\Upsilon(nS)$ dimuon mass region.
 
\subsection{Exclusive background}
\label{sec:exclusiveBkgd}
 
The Drell--Yan production of dimuons with a highly energetic photon, typically from final-state radiation,
$q\bar{q}\to\gamma^{*}/Z^{*}\to\mumu\gamma$,
constitutes a significant background contribution, exhibiting a characteristic resonant structure in the $m_{\mumu\gamma}$ distribution.
The shapes of this background in $m_{\mumu\gamma}$ and $m_{\mumu}$ are modelled using events simulated with \progname{Sherpa}\,2.2.10~\cite{Gleisberg:2008ta} at leading order with the NNPDF3.0 parton distribution function set in the $\psi(nS)$ and $\Upsilon(nS)$ $m_{\mumu}$ regions.
The normalisation of this background is determined from a fit to the data in the signal region following the selection described in Section~\ref{sec:selection}.
The resonant shape in $m_{\mumu\gamma}$ is modelled analytically as the sum of a Voigtian function and a threshold function defined as $f(x) = \sqrt{x - x_0}\,\mathrm{e}^{-A(x - x_0)}$, where
$A$ and $x_0$ are constants, $f(x) = 0$ for $x < x_0$, and $x$ is the three-body mass $m_{\mumu\gamma}$.
The Voigtian function describes the on-shell $Z$ production, which dominates in the $\Upsilon(nS)$ channels,
whereas the threshold function describes the off-shell $\gamma^*/Z^*$ production, which dominates in the $\psi(nS)$ channels. Different forms of the threshold function were considered, but the function above was found to provide the best description of the background shape.
The shape in $m_{\mumu}$ is non-resonant and is modelled with a first-order Chebyshev polynomial.
The parameters of the shape functions used to model each mass distribution are obtained from a fit to the simulated samples in each category once the signal region selection described in Section~\ref{sec:selection} is imposed.
The statistical uncertainty associated with the parameters of the $m_{\mumu\gamma}$ shape function is accounted for via nuisance parameters in the maximum-likelihood fit to data in Section~\ref{sec:results}, based on the fit to the simulated samples, and is found to have a negligible effect on the final result.

\subsection{Inclusive background}
\label{sec:inclusiveBkgd}
 
The dominant background arises from inclusive multi-jet or $\gamma$+jet events that involve the production of $\mathcal{Q}$ states, which subsequently decay into $\mumu$,
or the production of non-resonant dimuon pairs such as from the Drell--Yan process or from random combinations of muons; the photon candidate may be genuine or, typically, a misidentified jet.
The contribution from dimuon events in the inclusive background is separate from the exclusive $q\bar{q}\to\mumu\gamma$
background discussed in Section~\ref{sec:exclusiveBkgd},
as the latter involves a genuine photon candidate originating from the same hard-scattering process.
The complicated mixture of background contributions, which involve QCD processes and misidentification of physics objects, and the highly selective
phase-space region of interest makes it challenging to model this inclusive background accurately with simulation.
Furthermore, the features of the background shape, which exhibits a broad kinematic peak at the location of a possible $Z\to{\mathcal Q}\gamma$ signal, combined with the relatively low number of events in the signal region make background modelling through direct fits of parametric models to the data unsuitable.
For this reason, a generative approach is pursued, where the $m_{\mumu\gamma}$ shape of the inclusive background is obtained with a non-parametric data-driven model,
described in Ref.~\cite{Chisholm:2021pdn} and used in several previous analyses~\cite{HIGG-2014-03,HIGG-2016-23,HIGG-2016-05,HIGG-2016-13},
using templates to describe the kinematic distributions.
The background normalisation is extracted directly from a fit to the data,
and shape variations are incorporated in the background model in the final discriminating variable; these shape variations are also profiled in the fit.
 
The background model generation uses a sample of approximately $1.8\times 10^{4}$
$\psi(nS)\,\gamma$ and $8.9\times 10^{3}$
$\Upsilon(nS)\,\gamma$ candidate events. These events pass all  the
kinematic selection requirements described in Section~\ref{sec:selection}, except that the
$\mathcal{Q}$ and $\gamma$ candidates are not required to satisfy the
nominal isolation requirements, and a looser minimum $\pt^{\mumu}$ requirement of
$30~\GeV$ is imposed; these events define the background-dominated generation region
(GR).
The exclusive background contribution is subtracted from the data to prepare for the construction of the inclusive background model. The
exclusive background mass shape is obtained from a fit to the simulated events that meet the GR selection criteria. The background model is derived iteratively; in each iteration the
exclusive $q\bar{q}\to\mumu\gamma$
contribution and the background model are fitted to the data in the GR to derive the
exclusive background
normalisation.
PDFs are constructed to describe the distributions
of the relevant kinematic and isolation variables and their most important correlations.
The PDFs of these kinematic and isolation variables are
sampled to generate an ensemble of pseudocandidate events, each with
complete $\mathcal{Q}$ and $\gamma$ candidate four-vectors and their associated isolation values.
The important correlations among the kinematic and isolation variables of the background events,
in particular between the $\mathcal{Q}$ and $\gamma$ candidate transverse momenta $\pt^{\mathcal Q}$ and $\ptgamma$,
are retained in the generation of the pseudocandidate events through the
following sequential sampling scheme, shown diagrammatically in Figure~\ref{fig:BkgdModel}:
 
\begin{enumerate}
\item The $\mathcal{Q}$-candidate pseudorapidity $\eta_{\mathcal{Q}}$, mass $m_{\mathcal{Q}}$, and azimuthal angle $\phi_{\mathcal{Q}}$, are drawn independently from one-dimensional PDFs, and values for $\pt^{\mathcal{Q}}$ and $\pt^{\mathcal{\gamma}}$ are drawn simultaneously from a two-dimensional PDF.
\item The pseudorapidity difference between the $\mathcal{Q}$ and $\gamma$ candidates, $\Delta\eta(\mathcal{Q},\gamma)$, and the $\gamma$-candidate calorimeter isolation are drawn simultaneously from a three-dimensional PDF, based on the value previously drawn for $\pt^{\mathcal{Q}}$.
\item The $\mathcal{Q}$-candidate track isolation and the azimuthal angular separation between the $\mathcal{Q}$ and $\gamma$ candidates, $\Delta\phi(\mathcal{Q},\gamma)$, are drawn separately from two three-dimensional PDFs, given the selected values of the $\gamma$-candidate calorimeter isolation and $\pt^{\mathcal{Q}}$.
\item The $\gamma$-candidate track isolation is drawn from a three-dimensional PDF using the previously sampled values of the $\mathcal{Q}$-candidate track isolation and $\gamma$-candidate calorimeter isolation variables, and the values for $\eta_{\gamma}$ and $\phi_{\gamma}$ are calculated from the values drawn for $\eta_{\mathcal{Q}}$, $\phi_{\mathcal{Q}}$, $\Delta\eta(\mathcal{Q},\gamma)$, and $\Delta\phi(\mathcal{Q},\gamma)$.
\end{enumerate}
 
\begin{figure}[!htbp]
\begin{center}
\hspace{-0.50cm}
\includegraphics[width=1.0\textwidth]{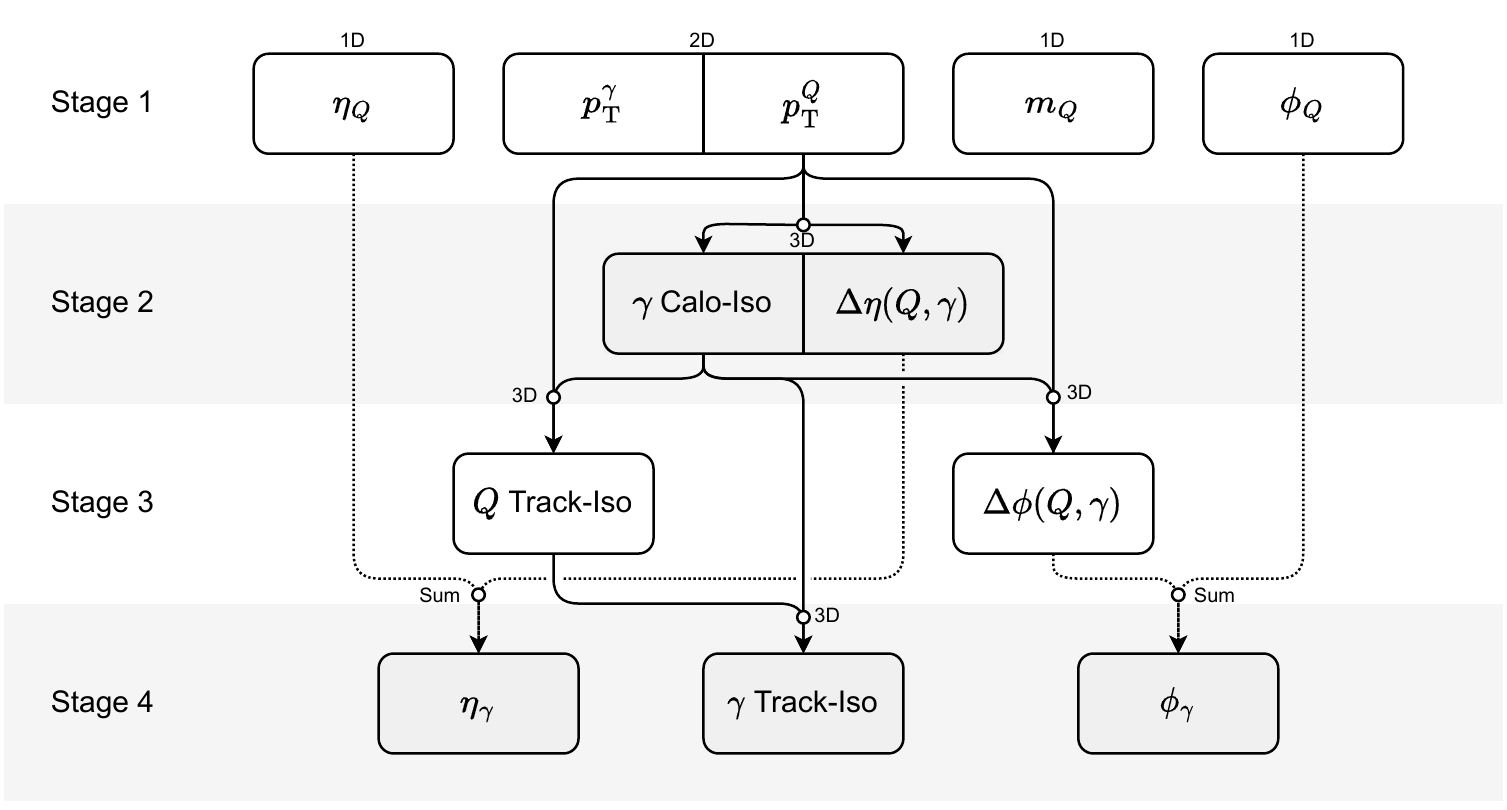}
\end{center}
\vspace{-0.6cm}
\caption{The data-driven sequential sampling method used to generate pseudocandidate events for the inclusive background model.
The labels `1D' and `2D' refer to the dimensionality of the PDFs used in the model generation of the variables underneath.
Vertices labelled `3D' signify that the output variable (or variables), indentifed by the arrow leading out of the vertex, is sampled from a three-dimensional PDF described in bins of the input variable (or variables), indentifed by the lines leading into the vertex.
If two variables share a border, they are sampled simultaneously from a joint PDF.
Vertices labelled `Sum' signify that the output variable is calculated directly from the sum of the input variables.\label{fig:BkgdModel}
}
\end{figure}
 
The nominal selection requirements are imposed and the
surviving pseudocandidates are used to construct templates for the
$m_{\mumu\,\gamma}$ distributions, which are then smoothed using a Gaussian kernel density
estimation method~\cite{Cranmer:2000du}.
Potential contamination of the GR sample from signal events is expected to be negligible.
Signal injection tests were performed where a significant amount of signal, much larger than the signal branching fraction excluded in the presented analysis, was added to the GR to investigate the effect of such a potential signal contamination on the model;
it was found that the presence of the signal is largely inconsequential to the shape of the background model and does not lead to any peaking structures in the background templates.
The ability to predict data using the background model is studied in several validation regions (VRs),
defined by selections looser than the nominal signal requirements and tighter than the generation region requirements.
The requirements imposed in each of the selection regions used in the analysis are summarised in Table~\ref{tab:regionDefinitions}.
The application of the $\pt^{\mumu}$ requirement (VR1), the muon isolation requirements (VR2), and the photon isolation requirement (VR3)
are each checked with this method, and in all regions the background model is found to describe the data well.
This is shown in the comparison of the background model with data in each VR in Figure~\ref{fig:BkgdVal},
which includes the exclusive background contribution.
The
exclusive background
shape in each selection region is obtained from a fit to the simulated events,
and the normalisation is extrapolated from the fit to data in the GR.

\begin{table}[htbp]
\centering
\caption{Summary of the selection regions used in the analysis. The term `Full' indicates the corresponding requirement applied in the SR, and discussed in Section~\ref{sec:selection}.
The relaxed photon isolation requires the sum of the $\pt$ of all ID tracks originating from the primary vertex and within $\Delta R = 0.2$ of the photon direction,
excluding any associated with the reconstructed photon, to be less than $20\%$ of $\ptgamma$ and
the sum of the $\pt$ of the calorimeter energy clusters within $\Delta R = 0.4$ of the photon direction,
excluding the energy of the reconstructed photon, to be less than $\left(2.45~\GeV + 0.4 \times \ptgamma~[\GeV]\right)$.
The relaxed $\mathcal{Q}$ isolation requires the sum of the $\pt$
of the ID tracks within $\Delta R = \min\{10~\GeV/(\ptmu~[\GeV]),
0.3\}$ of the leading muon to be less than 40\% of
$\pt^{\mathcal{Q}}$.\label{tab:regionDefinitions}
}
\vspace{0.2cm}
\begin{tabular}{|lr|lll|}
\hline
Region &  & \ptmumu                 & Photon Isolation  & ${\mathcal Q}$~Isolation \\ \hline
Generation Region   & (GR)      & $>30$~\GeV                     & Relaxed             & Relaxed \\
Validation Region 1 & (VR1)     & Full  & Relaxed                 & Relaxed \\
Validation Region 2 & (VR2)     & $>30$~\GeV                     & Relaxed             & Full \\
Validation Region 3 & (VR3)     & $>30$~\GeV                     & Full       & Relaxed \\
Signal Region       & (SR)      & Full  & Full  & Full \\ \hline
\end{tabular}
\end{table}

\begin{figure}[!htbp]
\begin{center}
\hspace{-0.50cm}
\subfigure[\texttt{$H(Z)\to\,\psi(nS)\gamma$ VR1}]{\label{fig:BkgdValJpsi_VR1}\includegraphics[width=0.34\textwidth]{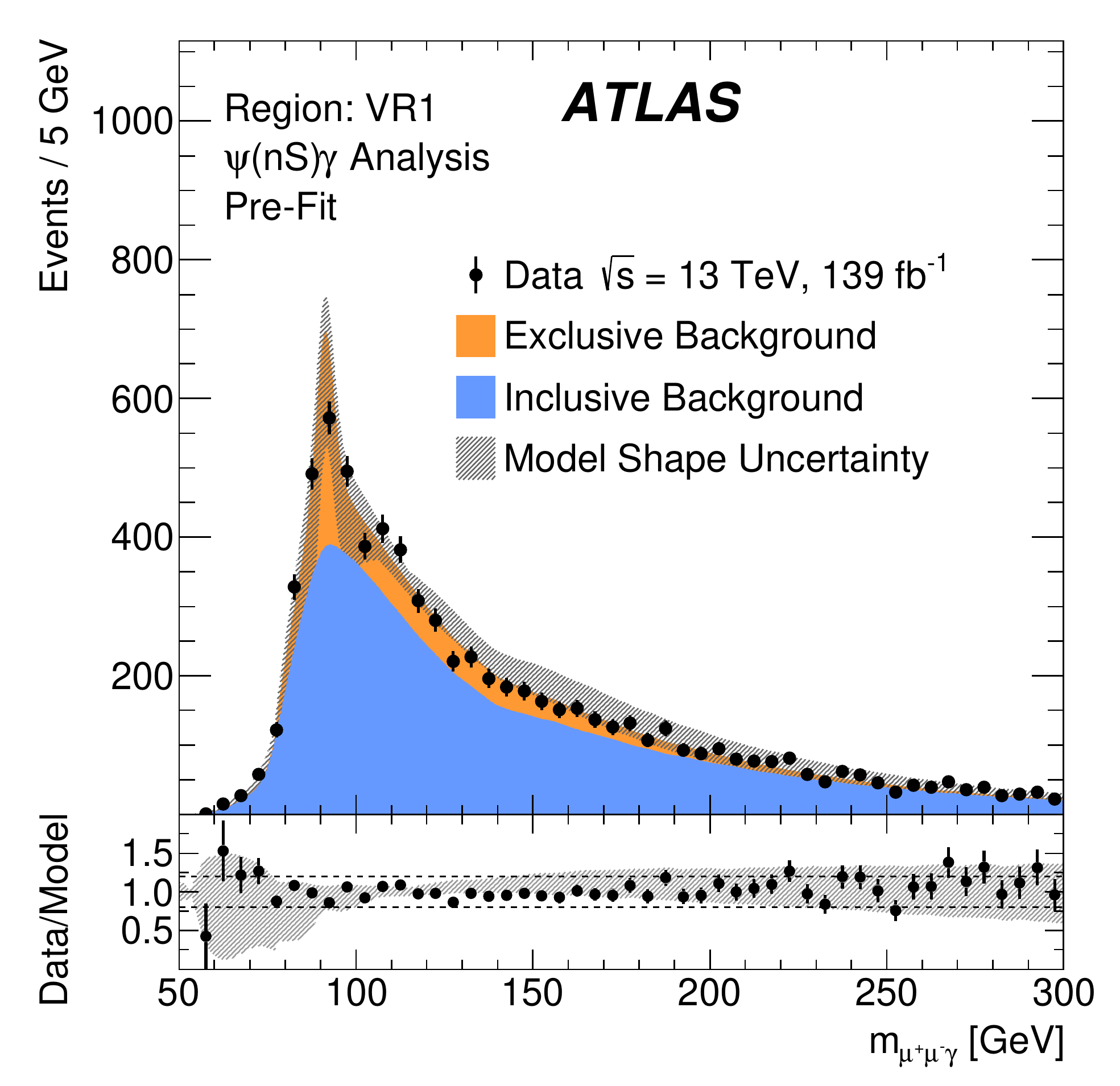}} \hspace{-0.30cm}
\subfigure[\texttt{$H(Z)\to\,\psi(nS)\gamma$ VR2}]{\label{fig:BkgdValJpsi_VR2}\includegraphics[width=0.34\textwidth]{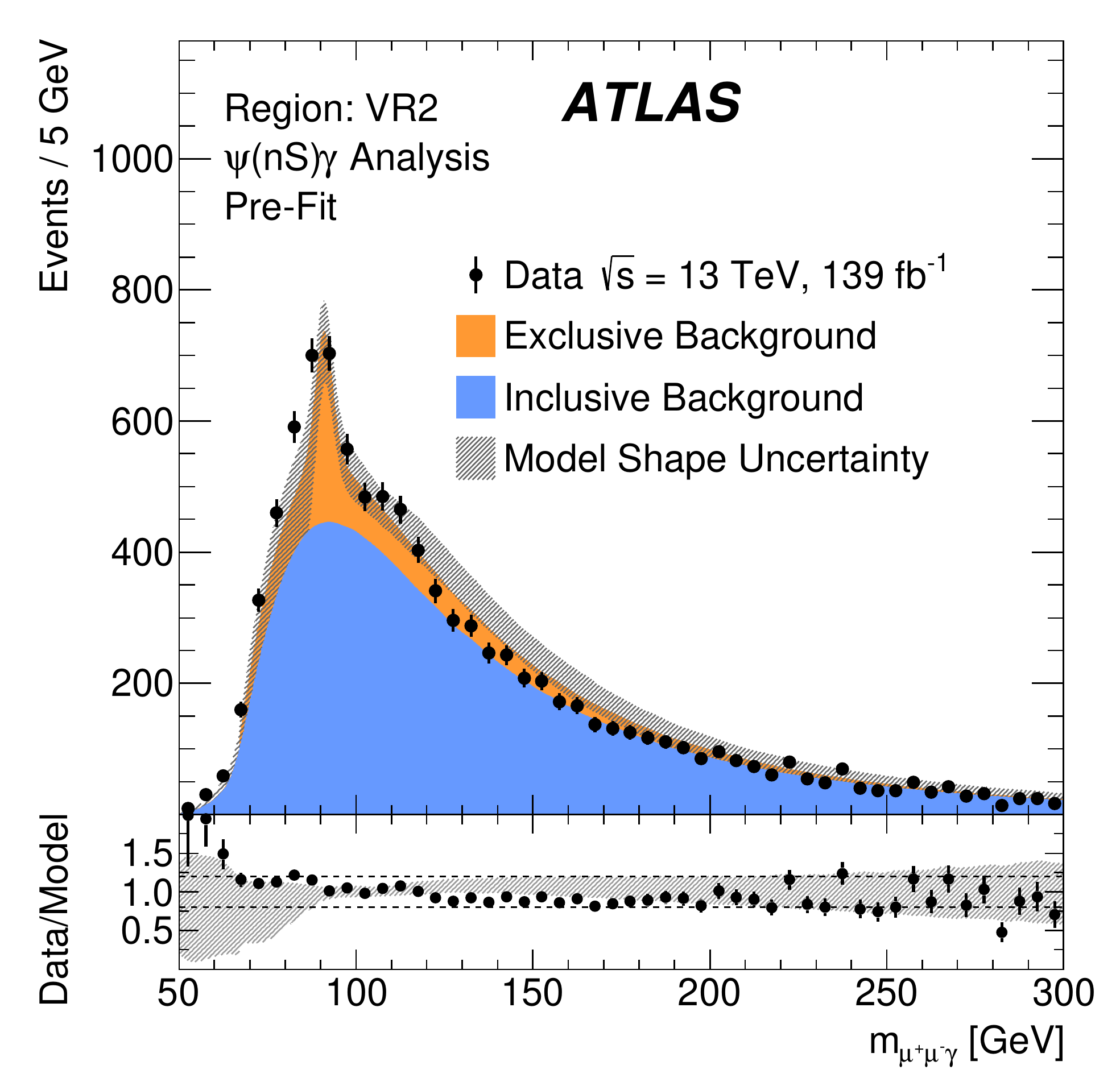}} \hspace{-0.30cm}
\subfigure[\texttt{$H(Z)\to\,\psi(nS)\gamma$ VR3}]{\label{fig:BkgdValJpsi_VR3}\includegraphics[width=0.34\textwidth]{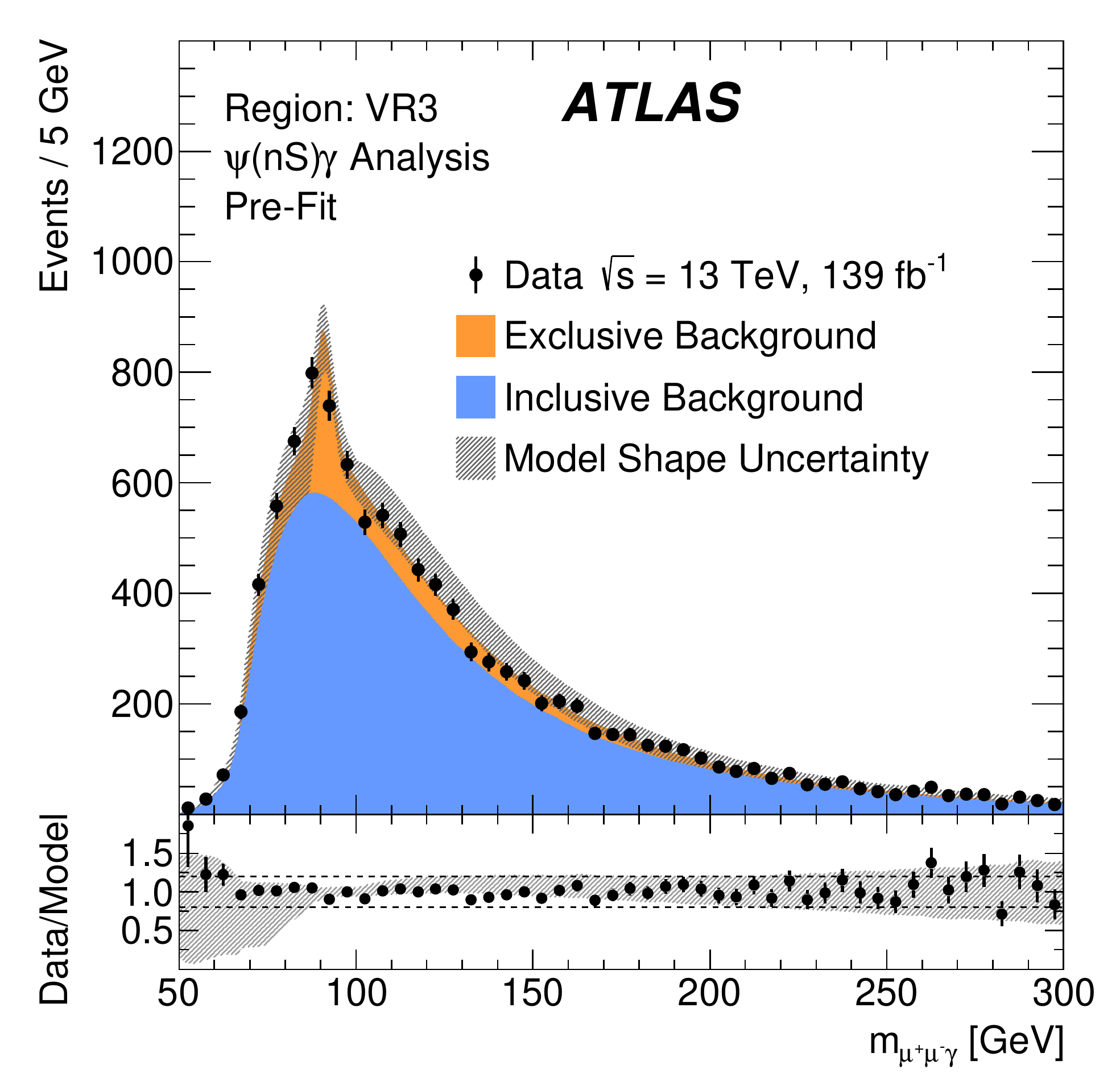}}
 
\hspace{-0.50cm}
\subfigure[\texttt{$H(Z)\to\,\Upsilon(nS)\gamma$ VR1}]{\label{fig:BkgdValUpsilon_VR1}\includegraphics[width=0.34\textwidth]{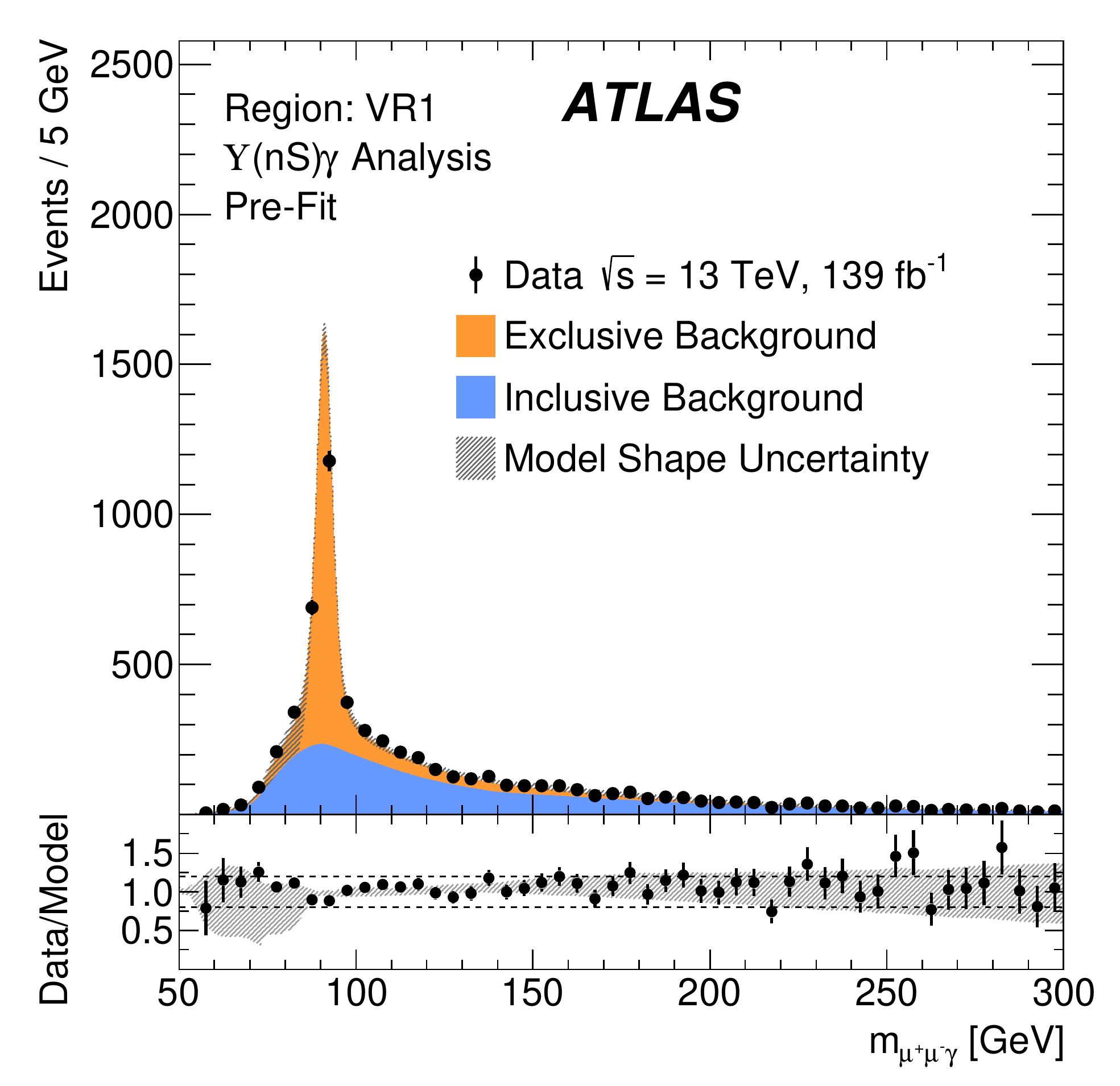}} \hspace{-0.30cm}
\subfigure[\texttt{$H(Z)\to\,\Upsilon(nS)\gamma$ VR2}]{\label{fig:BkgdValUpsilon_VR2}\includegraphics[width=0.34\textwidth]{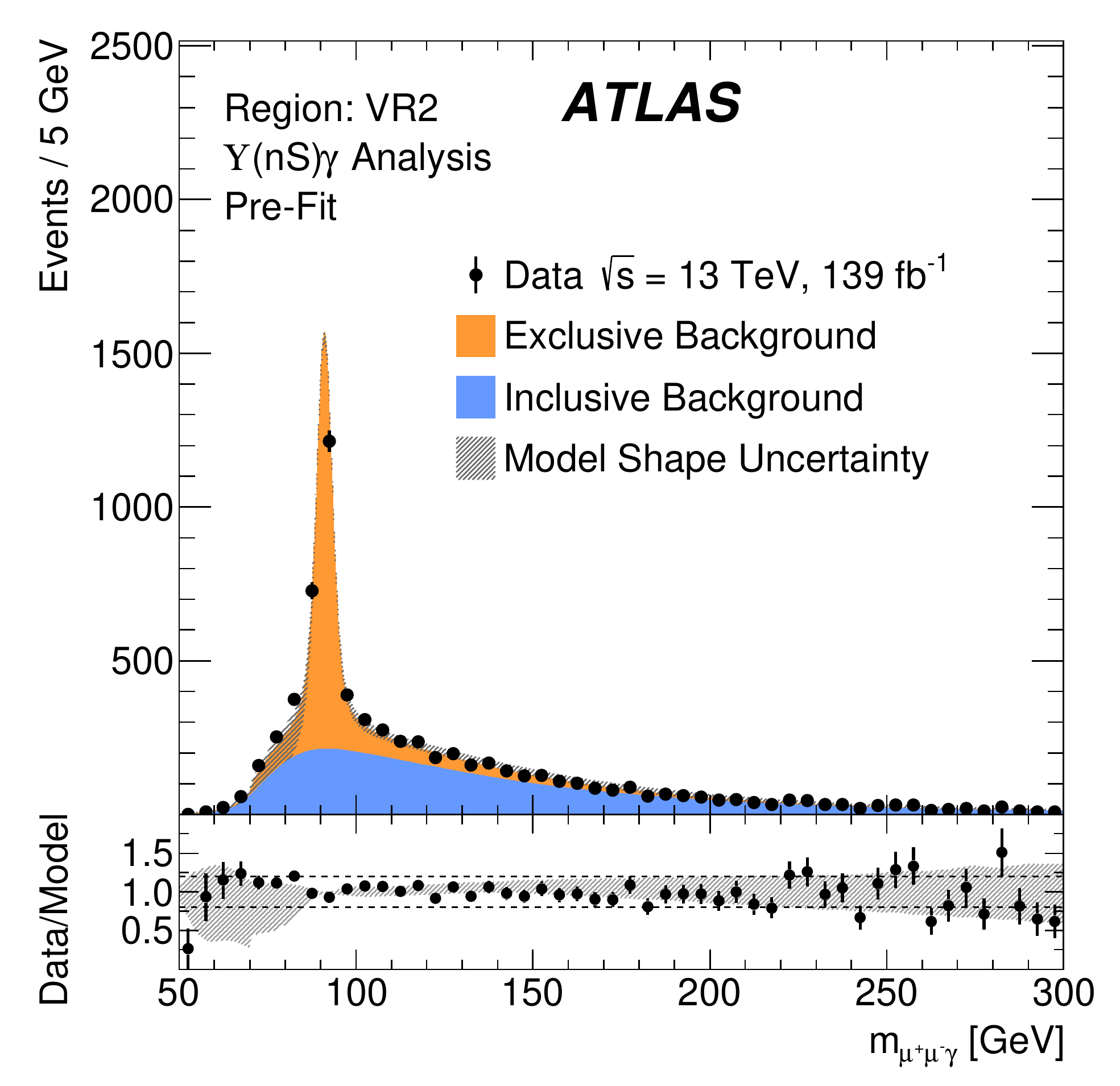}} \hspace{-0.30cm}
\subfigure[\texttt{$H(Z)\to\,\Upsilon(nS)\gamma$ VR3}]{\label{fig:BkgdValUpsilon_VR3}\includegraphics[width=0.34\textwidth]{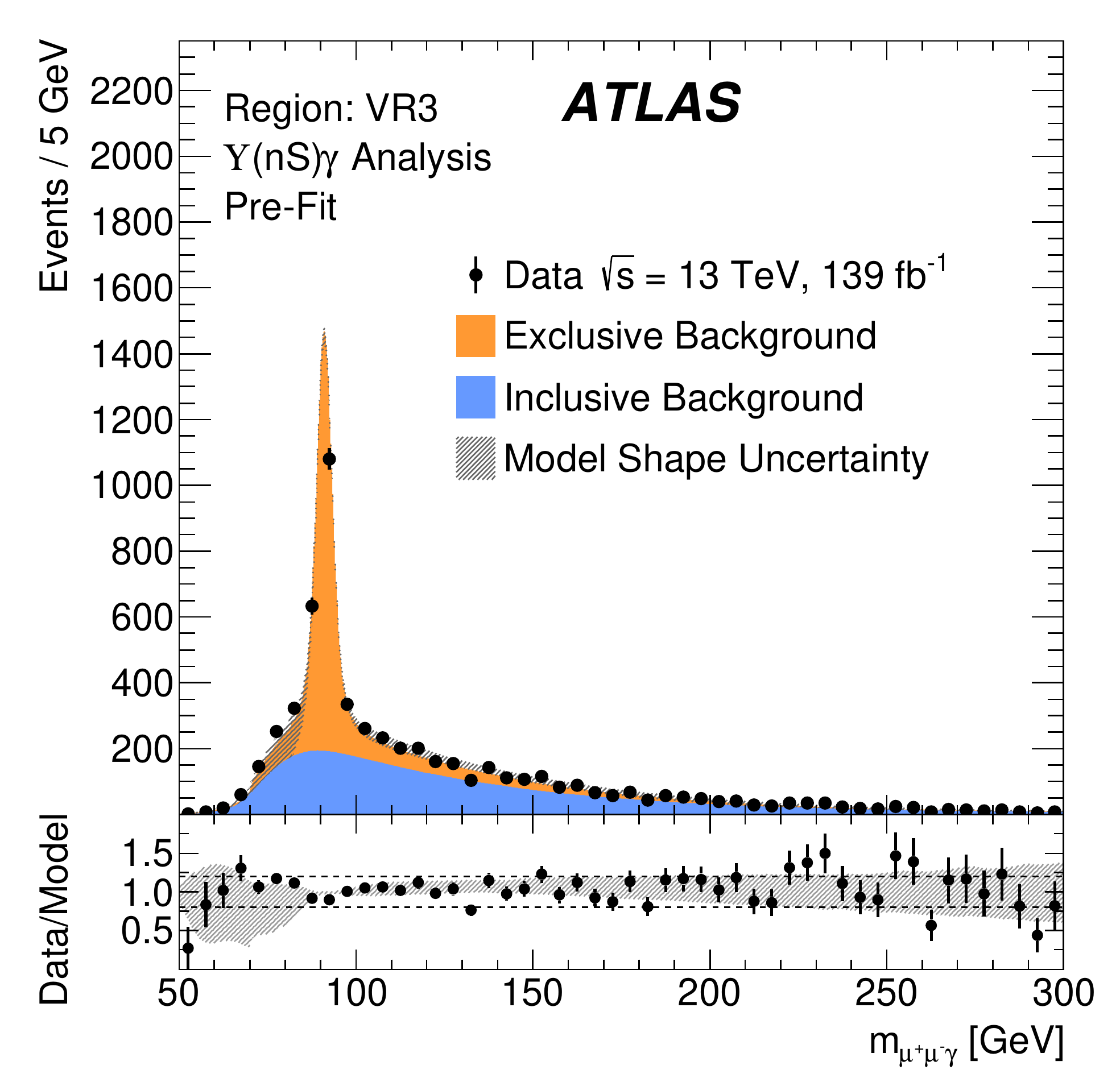}}
\end{center}
\vspace{-0.6cm}
\caption{The distribution of $m_{\mumu\gamma}$ in data compared to the prediction of the background model for (\subref{fig:BkgdValJpsi_VR1}, \subref{fig:BkgdValJpsi_VR2} and \subref{fig:BkgdValJpsi_VR3}) $H(Z)\to\psi(nS)\,\gamma$ and (\subref{fig:BkgdValUpsilon_VR1}, \subref{fig:BkgdValUpsilon_VR2} and \subref{fig:BkgdValUpsilon_VR3}) $H(Z)\to\Upsilon(nS)\,\gamma$ in the VR1, VR2 and VR3 validation regions.
The total background is normalised to the observed number of events within the region shown, where the ratio of the
exclusive and inclusive background components is extrapolated from the generation region. The uncertainty band corresponds to the uncertainty envelope derived from variations in the inclusive background modelling procedure. The dashed lines in the ratio plot in each figure indicate $1.2$ and $0.8$ on the $y$-axis. It should be noted that these plots are pre-fit, where the shapes of the inclusive and exclusive background components are fixed to the nominal template. In the maximum-likelihood fit to data in the signal region, the normalisation of each background is free, and their shapes are allowed to morph according to the defined shape variations.
\label{fig:BkgdVal}
}
\end{figure}
 
The shape of the background model in three-body mass, $m_{\mumu\gamma}$, is allowed to vary around the
nominal shape, and the parameters controlling these systematic variations
are treated as nuisance parameters in the maximum-likelihood fit described in Section \ref{sec:likelihood}.
Three such shape variations are implemented to allow the background template to adjust to the observed data.
The first variation is produced via a scale variation of the $\ptgamma$ distribution in the model,
which allows the peak of the three-body mass distribution to shift to lower or higher masses;
the second variation is produced via a linear distortion of the shape of the $\Delta\phi({\mathcal{Q}},\gamma)$ distribution,
which allows the width of the three-body mass distribution to increase or decrease;
the third variation is produced via a global tilt of the three-body mass distribution around a pivot point,
which allows the background to adapt to slopes with respect to the data.
The first two variations are straightforward alterations to the
underlying kinematics of the pseudocandidates, which cause
corresponding changes in the three-body mass.
Their nuisance parameters are loosely constrained by a Gaussian term in the likelihood, which arbitrarily assigns the $\pm1\sigma$ variations to a relatively large change in the shape, and these are subsequently constrained by the data.
The third variation is applied directly to the final three-body mass template and left unconstrained in the final fit.
 
The $m_{\mumu}$ distribution for $\psi(nS)\,\gamma$ candidates that meet the background model generation criteria is shown in Figure~\ref{fig:MassJpsiFitINC} and exhibits clear peaks at the $\Jpsi$ and $\psi(2S)$ masses.
In Figures~\ref{fig:MassUpsilonFitB} and  \ref{fig:MassUpsilonFitEC}, the corresponding distributions for the selected $\Upsilon(nS)\,\gamma$ candidates are shown respectively for the B and EC categories, where the individual $\Upsilon(1S)$,$\Upsilon(2S)$, and $\Upsilon(3S)$ peaks can be observed.
During the $m_{\mumu\gamma}$ background model generation, a value for $m_{\mumu}$ is drawn from the distributions shown in Figure~\ref{fig:MassFit}. No significant correlations between $m_{\mumu}$ and $m_{\mumu\gamma}$ are observed. Thus, the $m_{\mumu}$ distribution itself is modelled independently of $m_{\mumu\gamma}$ using analytical PDFs.
The $\psi(nS)$ and $\Upsilon(nS)$ peaks are modelled with Gaussian PDFs, while the inclusive background is modelled with a first-order Chebyshev polynomial function; these functions were found to be sufficient to describe the data, given its statistical uncertainty.
The parameters of these PDFs are obtained by means of a one-dimensional fit to events selected in the background model generation region, as shown in Figure~\ref{fig:MassFit}, and are used to model the $m_{\mumu}$ distribution in the maximum-likelihood fit in Section~\ref{sec:likelihood}. It is noted that the slope of the inclusive background is allowed to adapt to the observed data in the fit to the signal region, and that the relative normalisations of the different contributions to $m_{\mumu}$ are left free.
 
\begin{figure}[h]
\centering
\subfigure[]{\includegraphics[width=0.32\linewidth]{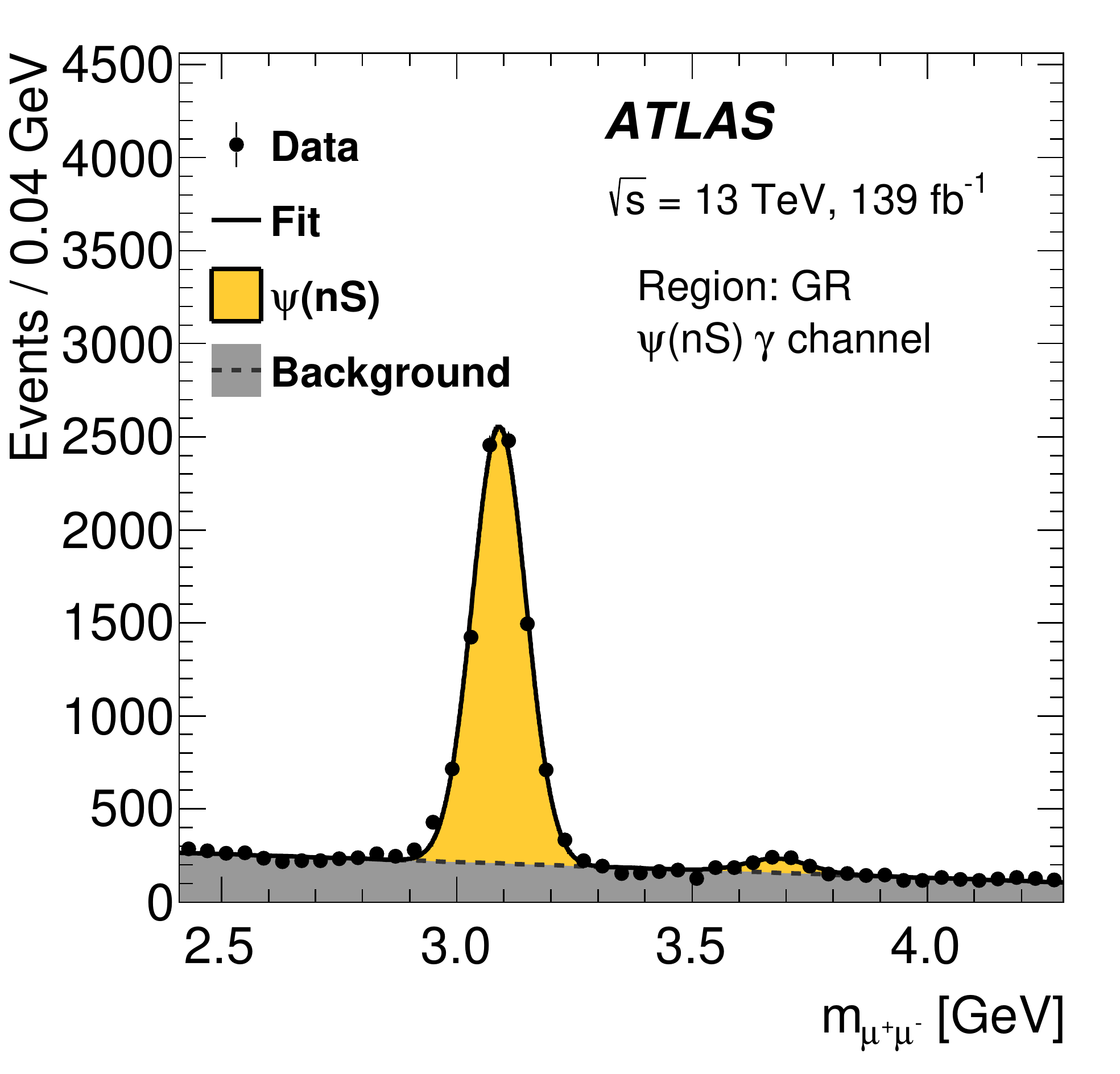}\label{fig:MassJpsiFitINC}}
\subfigure[]{\includegraphics[width=0.32\linewidth]{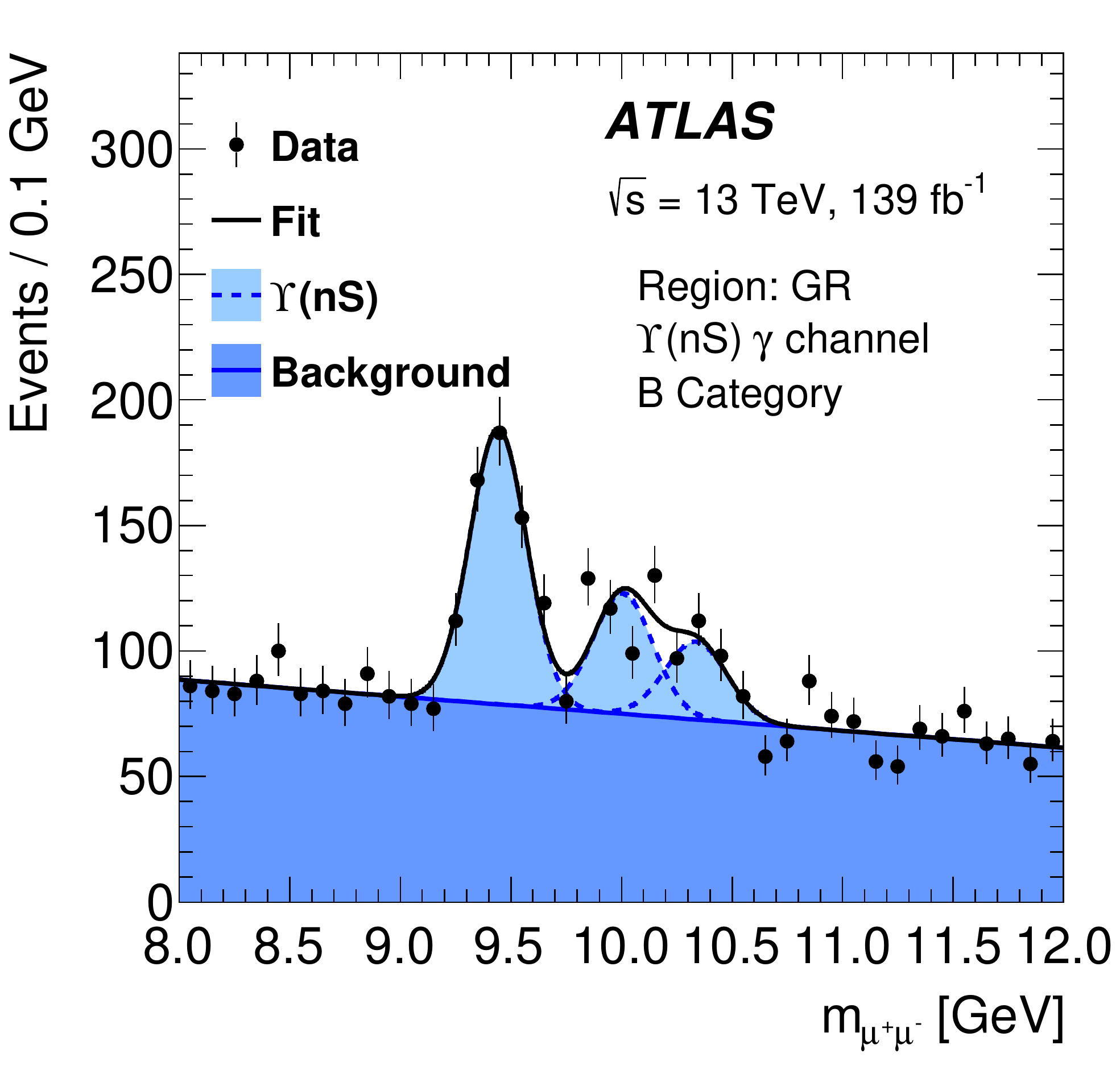}\label{fig:MassUpsilonFitB}}
\subfigure[]{\includegraphics[width=0.32\linewidth]{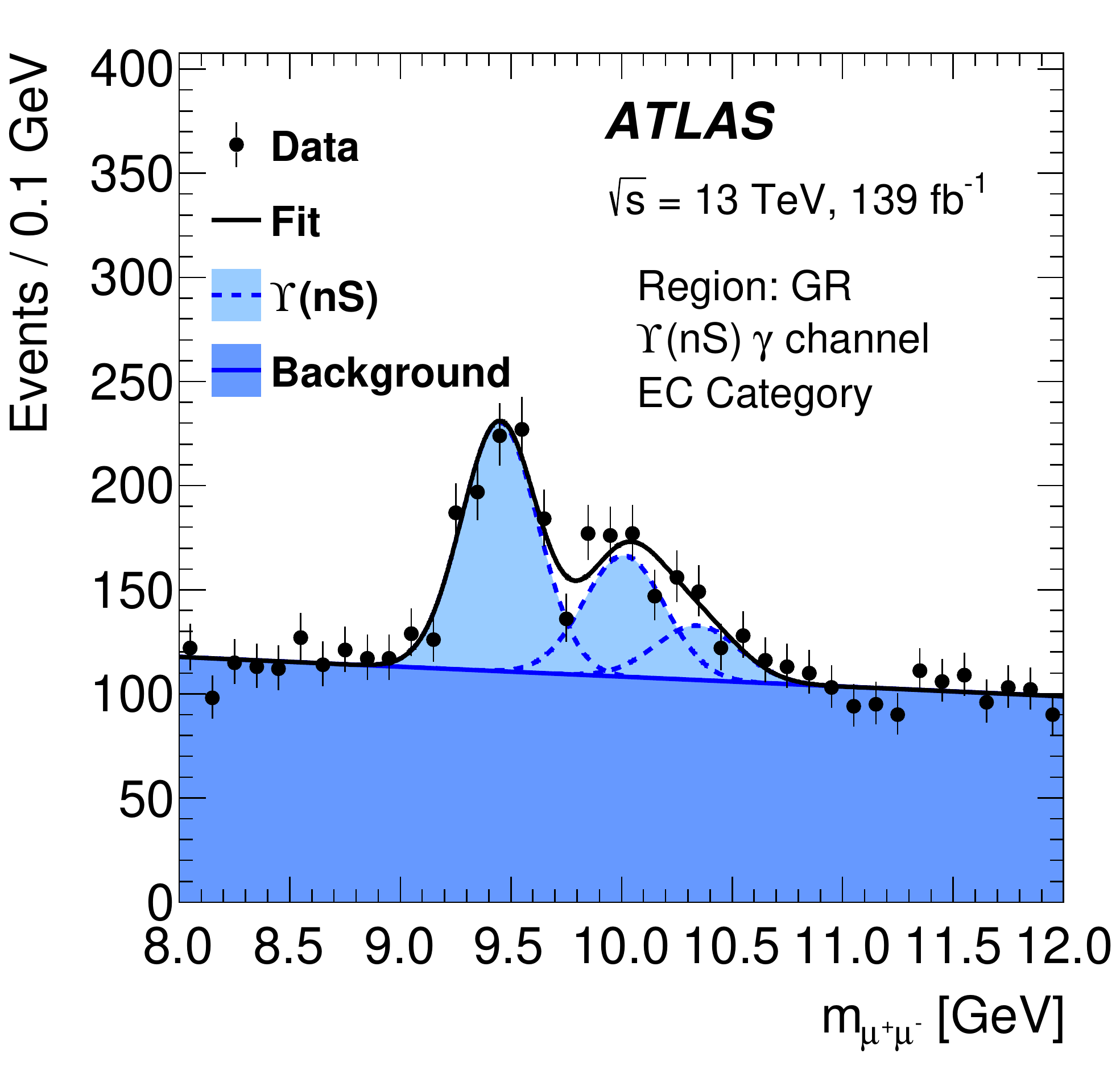}\label{fig:MassUpsilonFitEC}}
\vspace{-0.1cm}
\caption{Distribution of $\mumu$ invariant mass for \subref{fig:MassJpsiFitINC} $\psi(nS)\,\gamma$,
\subref{fig:MassUpsilonFitB} $\Upsilon(nS)\,\gamma$ barrel category (B) and \subref{fig:MassUpsilonFitEC} $\Upsilon(nS)\,\gamma$ endcap category (EC). Candidates satisfy the requirements of the
background generation region (GR) defined in Section~\ref{sec:inclusiveBkgd}. The error bars on the data points denote their statistical uncertainty.
\label{fig:MassFit}}
\end{figure}


\section{Statistical methods}
\label{sec:likelihood}
 
The data selected by the signal region criteria are compared with background and signal predictions using a two-dimensional (2D) unbinned maximum-likelihood fit in $m_{\mumu\,\gamma}$ and $m_{\mumu}$, for events with $m_{\mumu\gamma}<300~\GeV$.
This allows the $H(Z)\rightarrow\mathcal{Q}\,\gamma$ signals to be distinguished from each other, as well as the from the exclusive and inclusive background contributions.
The likelihood function ${\cal L}$ for each of the $\psi(nS)$ and $\Upsilon(nS)$ analyses is constructed using the signal and background models
described in Sections~\ref{sec:signalmodel} and~\ref{sec:backgroundmodel}.
The parameters of interest $\vec{\mu}=\{\mu_i\}$ are
the signal strengths,
which correspond to the signal rate
normalised to the SM expectation,
for each of the Higgs and $Z$ boson
signals counted by the index $i$. The likelihood for the $\psi(nS)$ analysis when $n$ events are observed in the signal region is defined as:\footnote{The likelihood for the $\Upsilon(nS)$ analysis is defined similarly, but with the events divided further into two exclusive categories.}
 
\begin{equation*}
\begin{split}{\cal L}(\vec{\mu},\vec{b}, \alpha,\theta,\theta') = & {\cal P}(n|\sum^{4}_{i=1}\mu_{i}\cdot s_{i} (\vec{\alpha})+\sum^{4}_{j=1} b_j) \times \prod_{r} {\cal G}(\alpha_r|0,1) \\
& \times \prod_{k=1}^{n}\left(\sum^{4}_{i=1}{\cal F}^{s}_{i}\cdot{\cal S}_i(m^k_{\mumu\gamma},m^k_\mumu|\mu_i)+ \sum^{4}_{j=1}{\cal F}^{b}_{j}\cdot{\cal R}_j(m^k_{\mumu\gamma}|\vec{\theta}){\cal M}_j(m^k_\mumu|\theta')\right)\\
& \times \prod_{l} {\cal G}(\theta_l|0,1) \\
\end{split}
\end{equation*}

In the above equation, ${\cal P}$ is the Poisson distribution for $n$ observed events given the total signal and background. The symbol $s_i(\vec{\alpha})$ denotes the expected SM signal yield for signal $i$ as modified by the nuisance parameters $\vec{\alpha}$, which correspond to the signal normalisation systematic uncertainties discussed in Section~\ref{sec:signalmodel}. These nuisance parameters are counted by index $r$ and are constrained with standard Gaussian PDF terms, ${\cal G}$.
The normalisation parameters associated with each independent background contribution are denoted by $\vec{b}=\{b_j\}$. These are not constrained and are determined directly from the fit to the data.
In the case of the $\psi(nS)$ analysis, the background components are
exclusive $q\bar{q}\to \mumu\gamma$ production, inclusive $J/\psi$ decays, inclusive $\psi(2S)$ decays, and inclusive non-resonant dimuon production.\footnote{In the case of the $\Upsilon(nS)$ analysis, the background components are
exclusive $q\bar{q}\to \mumu\gamma$ production, inclusive $\Upsilon(1S)$, $\Upsilon(2S)$, and $\Upsilon(3S)$ decays, and inclusive non-resonant dimuon production, and are fit separately in the B and EC categories.}
The symbols ${\cal F}^s_i$ and ${\cal F}^b_j$ denote the signal $i$ and background $j$ as fractions of the total signal and background, respectively.
The shape of signal $i$ in $(m_{\mumu\gamma},m_\mumu)$ is given by the PDF ${\cal S}_i$. Correspondingly, the shape of the background component $j$ is given by the product of PDFs ${\cal R}_j(m_{\mumu\gamma})$ and ${\cal M}_j(m_\mumu)$, since no correlations between $m_\mumu$ and $m_{\mumu\gamma}$ are observed in background shape. The nuisance parameters $\vec{\theta}$ parameterise the systematic variations of the background shapes discussed in Section~\ref{sec:backgroundmodel}. They are counted by index $l$ and are constrained with standard Gaussian terms.
The only exception is the nuisance parameter related to the `global tilt' systematic shape variation of the inclusive background, which is not constrained.
The nuisance parameter $\theta'$ corresponds to the slope in $m_\mumu$ of the non-resonant component of the inclusive background, which is free in the fit.
 
Upper limits are set on the branching fractions for the Higgs and $Z$~boson decays into
$\cal{Q}\,\gamma$ using the CL$_\mathrm{s}$ modified frequentist formalism~\cite{cls}
with the profile-likelihood-ratio test statistic and the asymptotic approximations
derived in Ref.~\cite{Cowan:2010js}.
When setting limits for one of the signals, the other potential signal contributions are treated as nuisance parameters and are profiled in the fit.
Only the decays $\mathcal{Q}\rightarrow\mumu$ are considered in the limit setting, feed-down from $Q'\rightarrow Q + X$ transitions are not included.
 
The ability of the fit to identify potential signal contributions was verified through signal injection tests.
Signals corresponding to a branching fraction of $5\times 10^{-4}$ and $5\times 10^{-7}$ for the Higgs and $Z$ boson decays, respectively, were injected into
an Asimov dataset~\cite{Cowan:2010js} constructed from a background-only fit in the signal region.
The full fit accurately recovered the injected signals.


\section{Results}
\label{sec:results}
 
In total, 3394 events are observed in the $\psi(nS)\,\gamma$ signal region and 3577 events are observed in the $\Upsilon(nS)\,\gamma$ signal region.
The results of the background-only fits for the $\psi(nS)\,\gamma$ and $\Upsilon(nS)\,\gamma$ analyses are shown in
Figures~\ref{fig:psiFits} and~\ref{fig:upsilonFits} respectively, where the signal distributions shown correspond to the extracted 95\% confidence-level (CL) branching fraction upper limits.
The expected and observed numbers of background events within the $m_{\mumu\,\gamma}$ ranges relevant to the Higgs and $Z$ boson signals are given in Table~\ref{tab:events},
where the expected backgrounds are obtained from these background-only fits.
The exclusive contribution to the total background in these regions of relevance ranges from approximately 10\% for the $H\rightarrow J/\psi\,\gamma$ and 22\% for $H\rightarrow \psi(2S)\,\gamma$
to 21\% and 41\% for the corresponding $Z$ boson decay searches. For the $\Upsilon(nS)\,\gamma$ analysis the
exclusive contribution ranges from 24\% to 29\% for the ranges of relevance for the Higgs boson searches, while for the ranges of relevance for the $Z$ boson searches it is between 75\% and 79\%. Table~\ref{tab:events} also shows the expected number of signal events for reference branching fractions near the sensitivity of the analysis: $10^{-3}$ for $H\rightarrow\mathcal{Q}\,\gamma$ and $10^{-6}$ for $Z\rightarrow\mathcal{Q}\,\gamma$.
 
\begin{figure}[!htbp]
\centering
\subfigure[]{\includegraphics[width=0.45\linewidth]{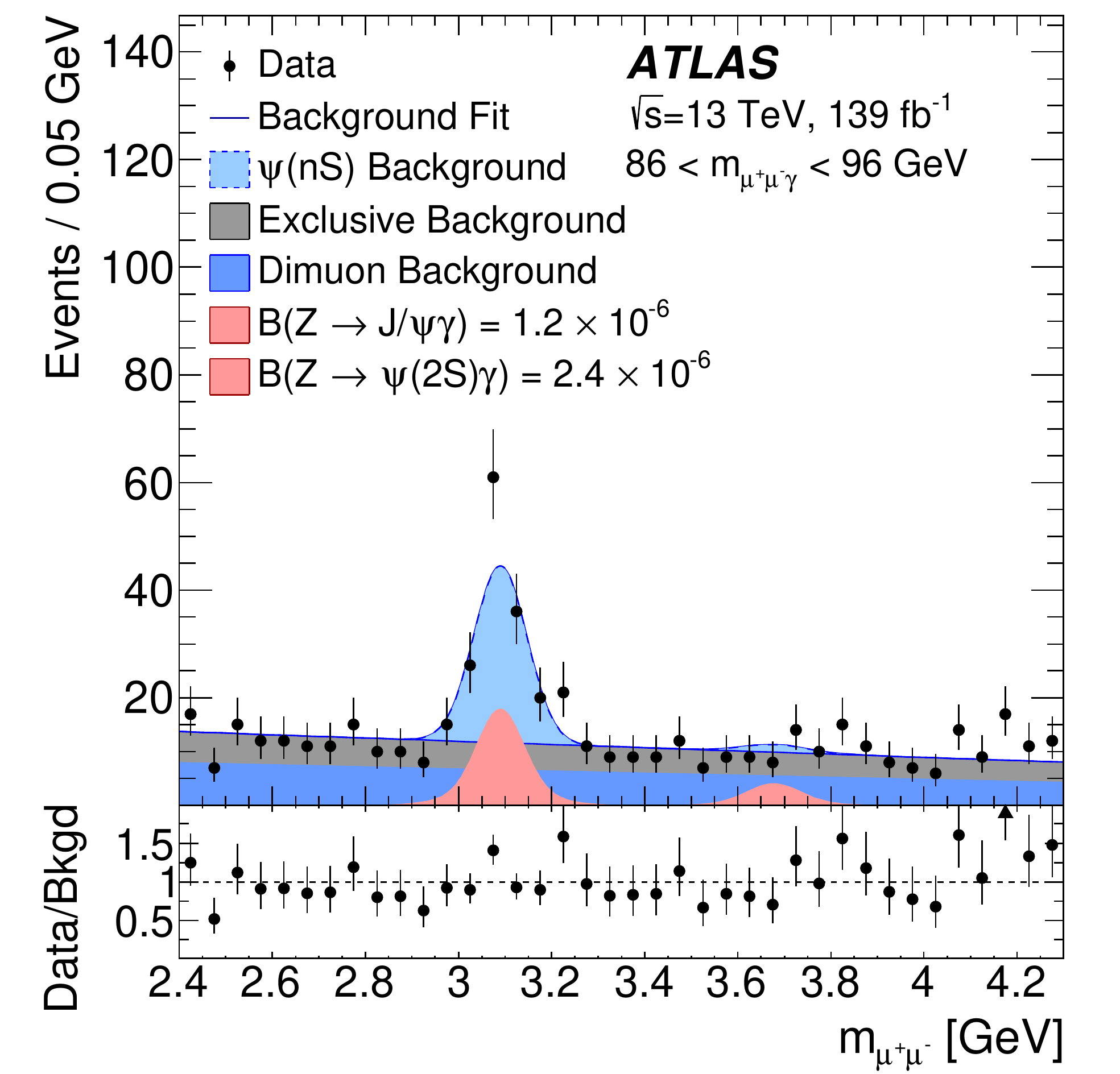}\label{fig:psi_mumu_ZCut}}
\subfigure[]{\includegraphics[width=0.45\linewidth]{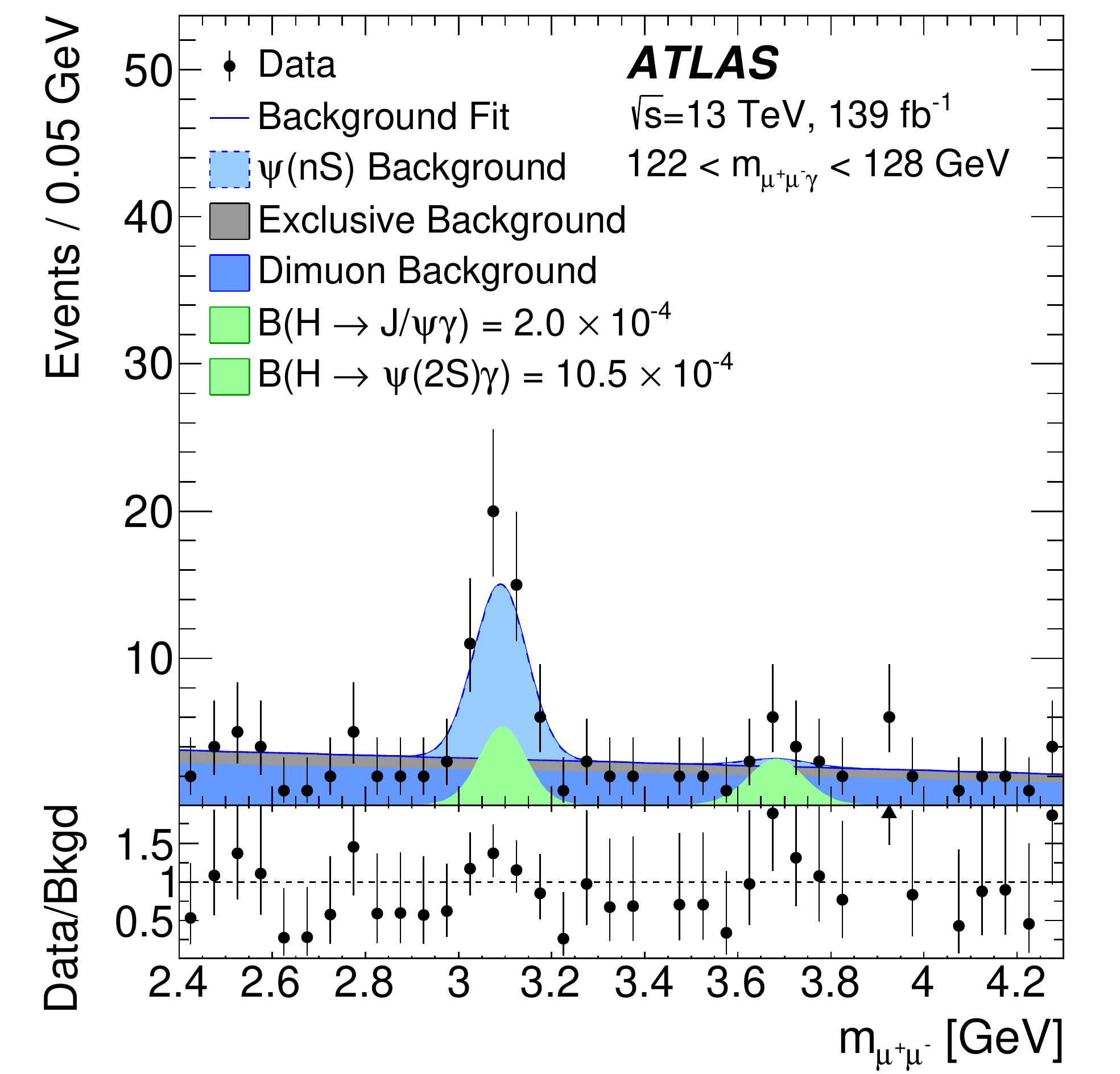}\label{fig:psi_mumu_HCut}}
\subfigure[]{\includegraphics[width=0.45\linewidth]{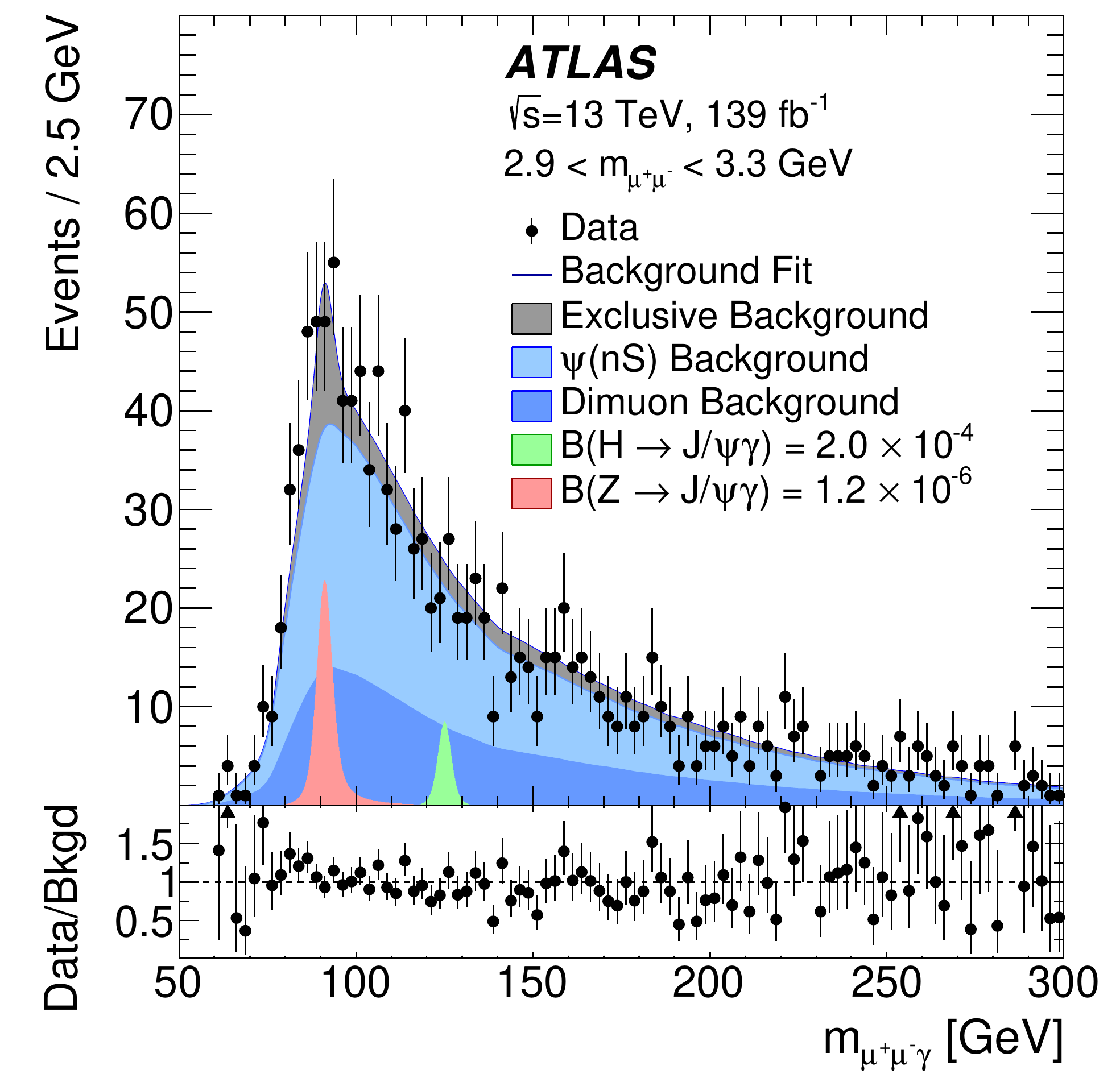}\label{fig:psi_mumugamma_1SCut}}
\subfigure[]{\includegraphics[width=0.45\linewidth]{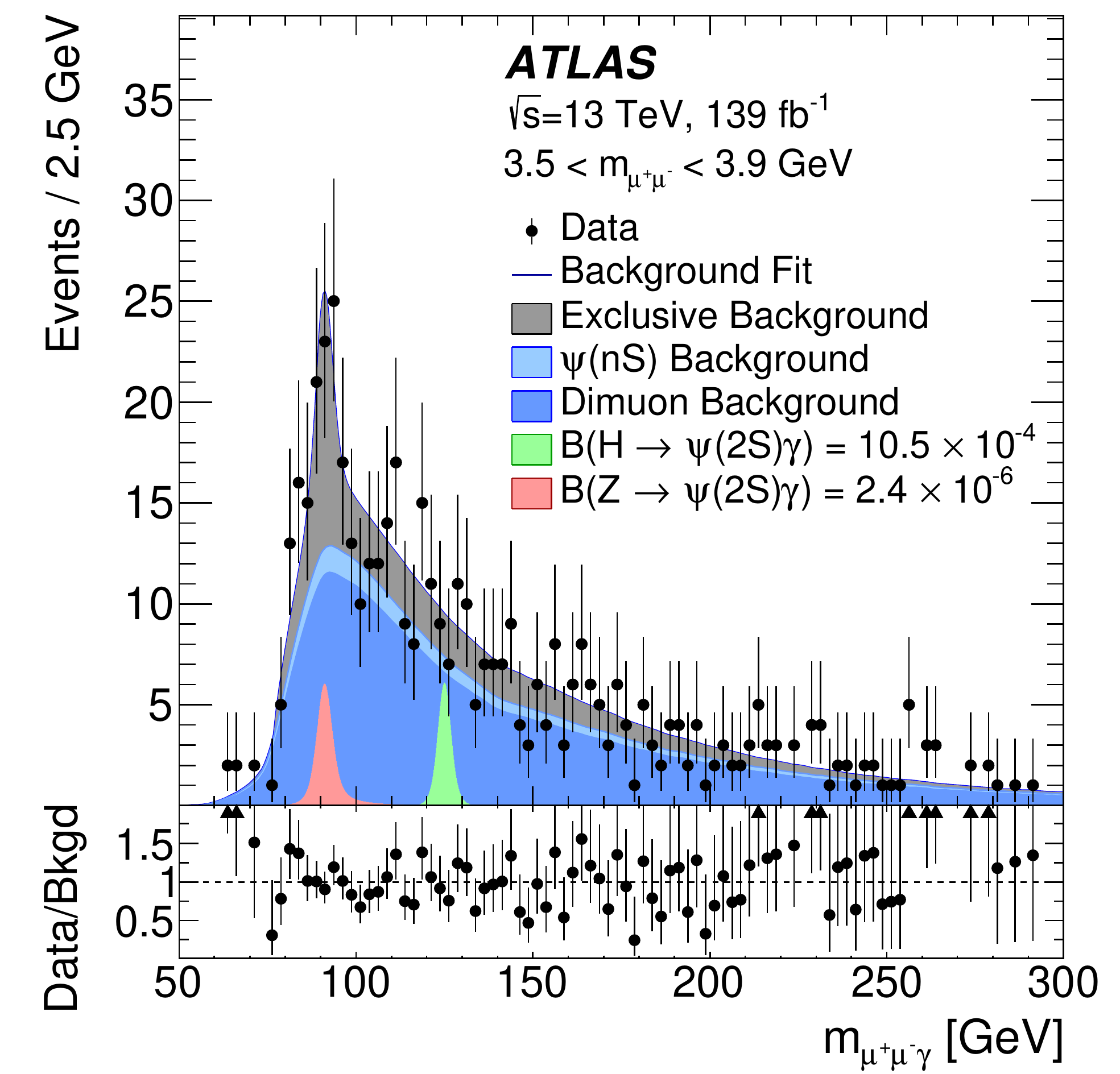}\label{fig:psi_mumugamma_2SCut}}
\caption{Projection of the $\psi(nS)$ channel fit in $m_{\mumu}$ for the~\subref{fig:psi_mumu_ZCut} $Z$ boson and~\subref{fig:psi_mumu_HCut} Higgs boson $m_{\mumu\gamma}$ regions, and in $m_{\mumu\gamma}$ for the~\subref{fig:psi_mumugamma_1SCut} $\jpsi$ and~\subref{fig:psi_mumugamma_2SCut} $\psi(2S)$ $m_{\mumu}$ regions. The dimuon/${\cal Q}(nS)$ backgrounds in the legend refer to the inclusive background contribution which is non-resonant/resonant in $m_\mumu$. The branching fraction of each of the signals is set to the observed $95\%$ CL upper limit.\label{fig:psiFits}}
\end{figure}
 
\begin{figure}[!htbp]
\centering
\subfigure[]{\includegraphics[width=0.45\linewidth]{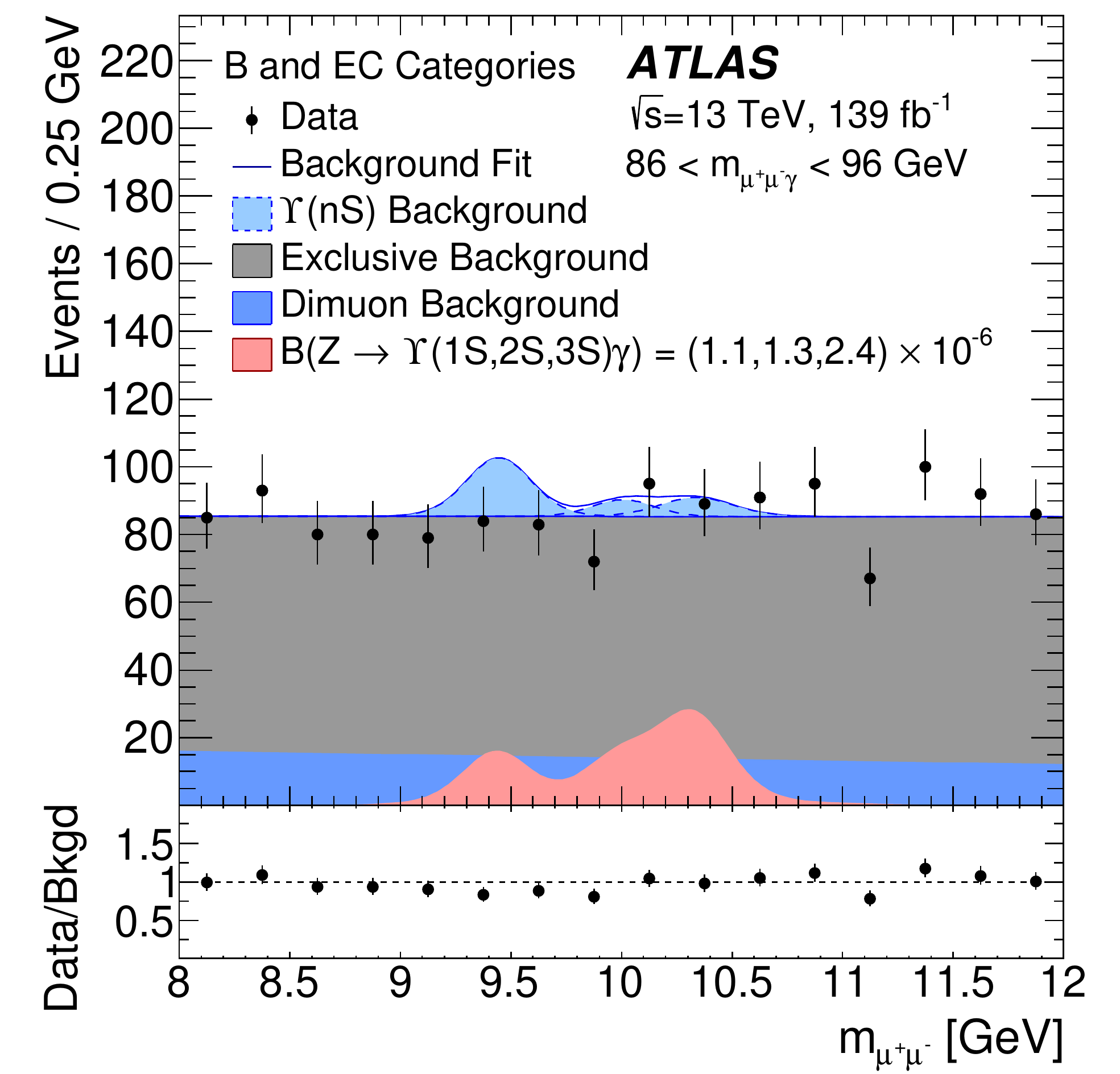}\label{fig:upsi_mumu_ZCut}}
\subfigure[]{\includegraphics[width=0.45\linewidth]{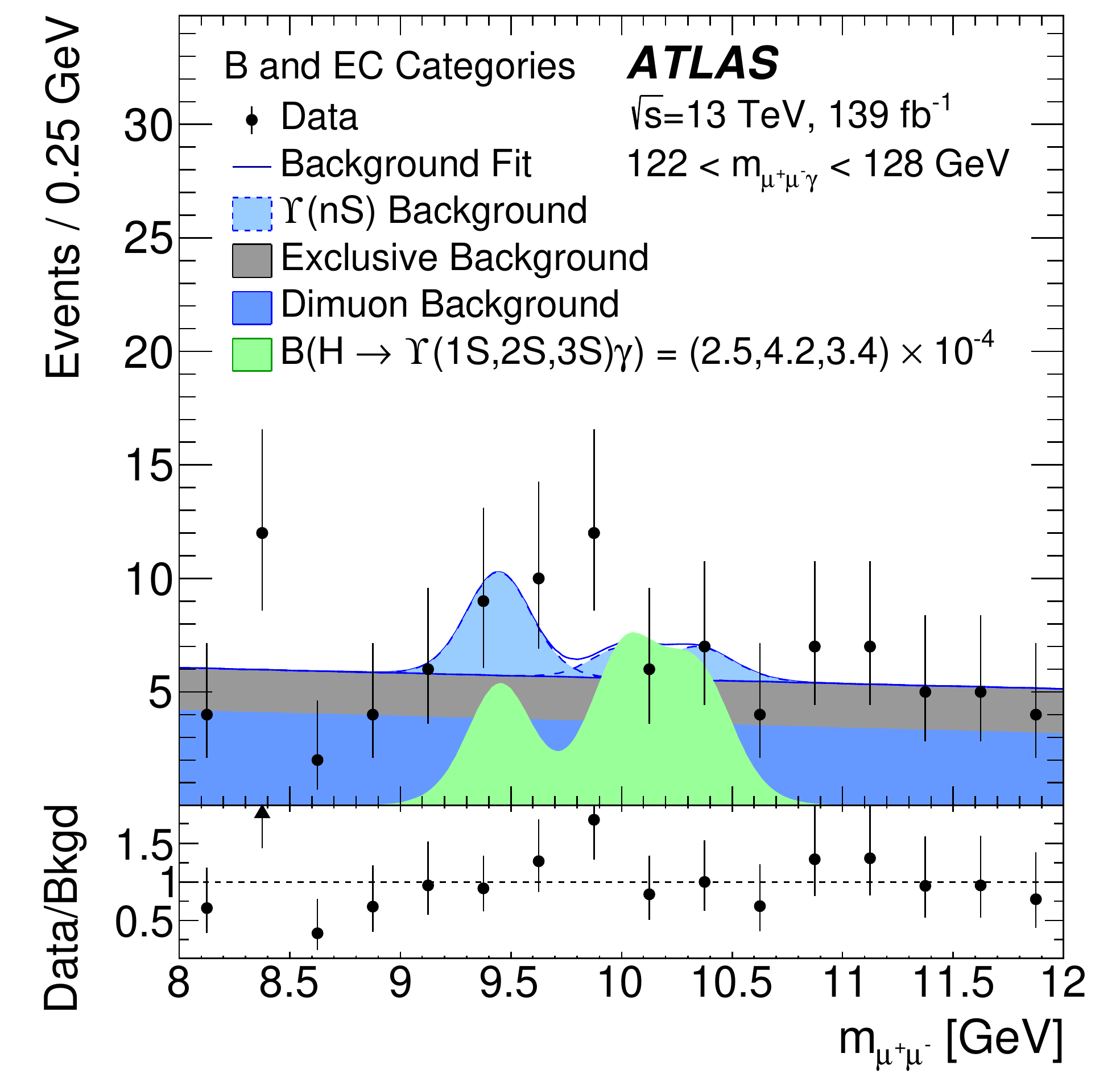}\label{fig:upsi_mumu_HCut}}
\subfigure[]{\includegraphics[width=0.45\linewidth]{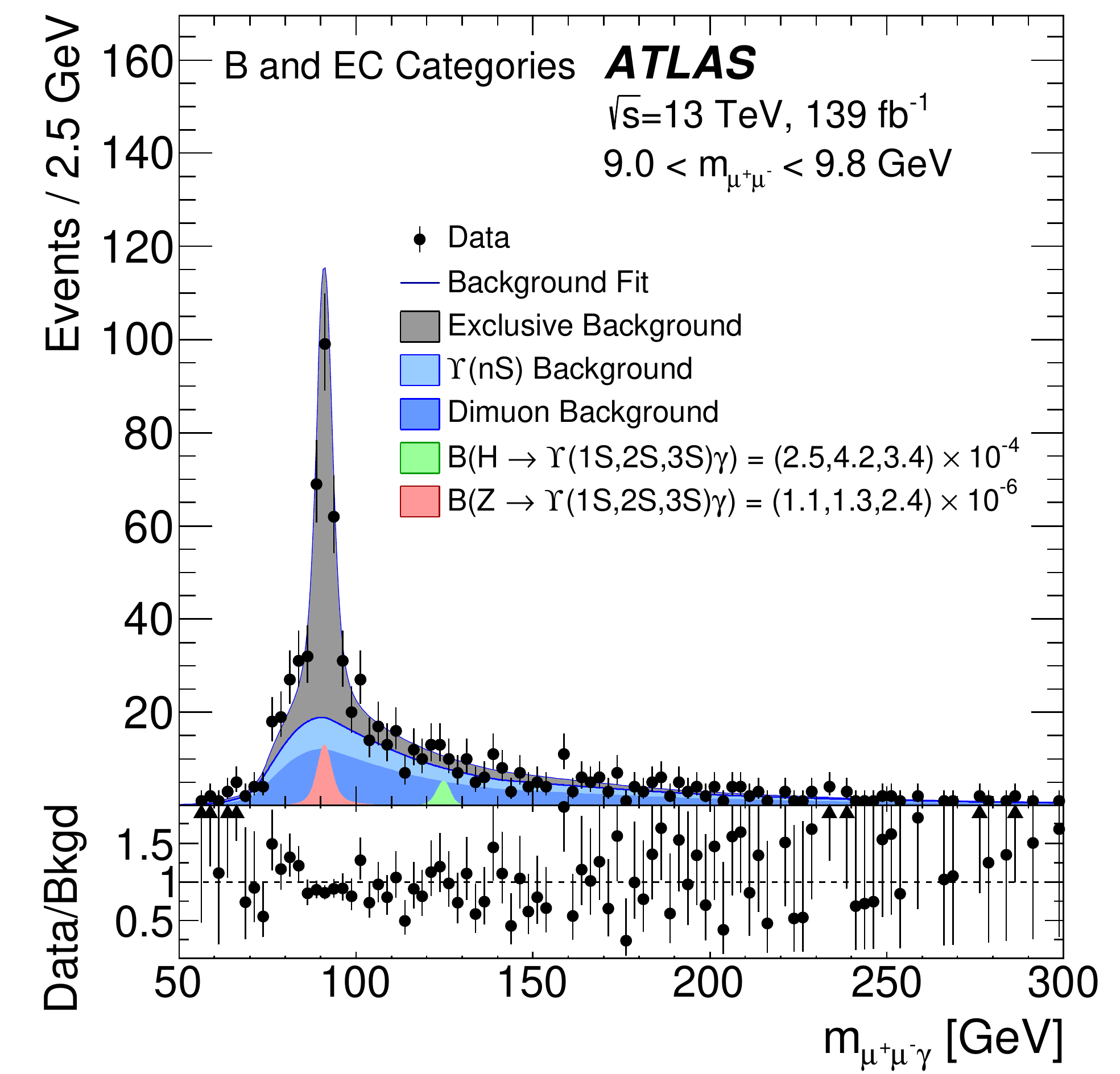}\label{fig:upsi_mumugamma_1SCut}}
\subfigure[]{\includegraphics[width=0.45\linewidth]{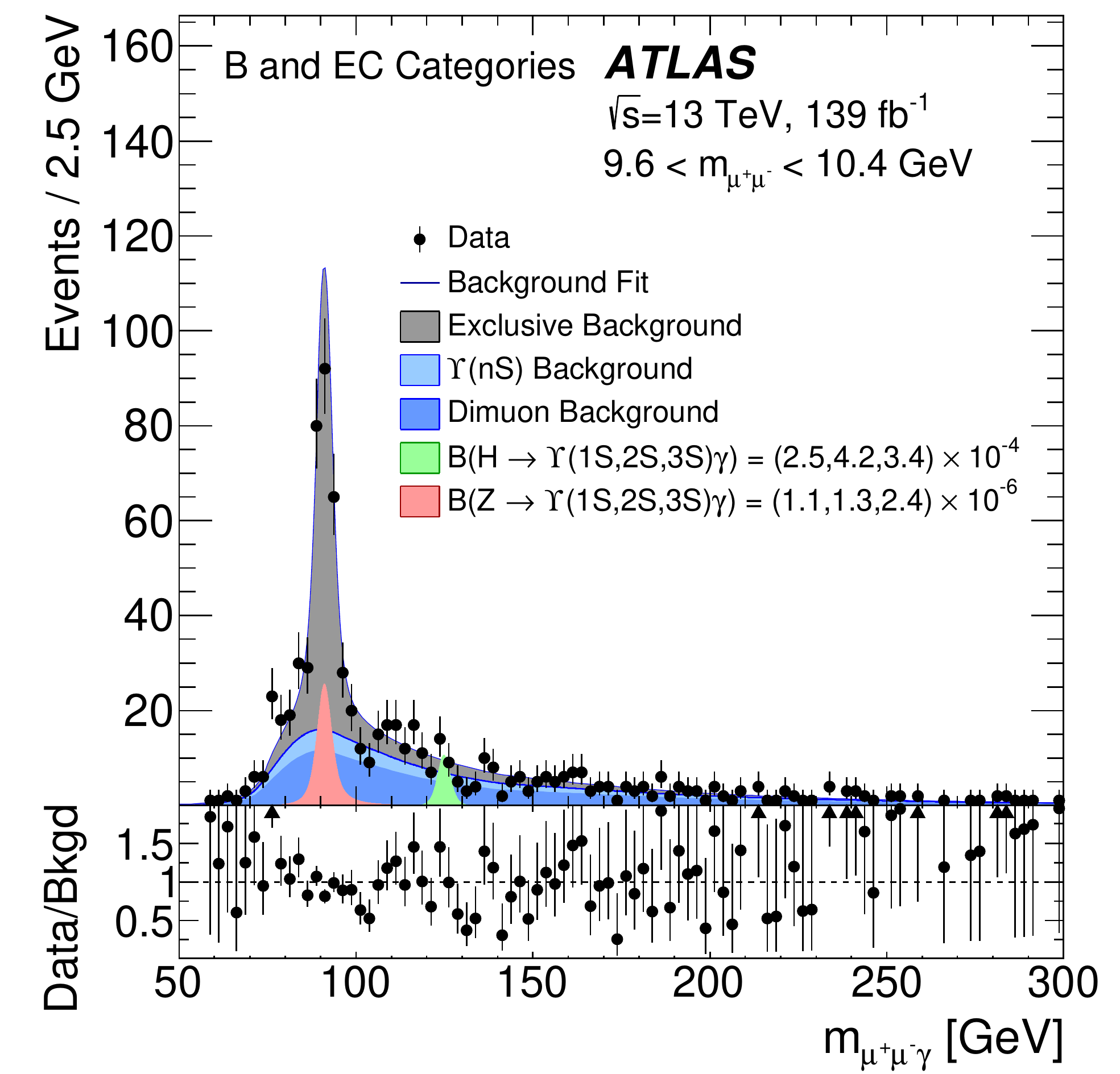}\label{fig:upsi_mumugamma_2SCut}}
\subfigure[]{\includegraphics[width=0.45\linewidth]{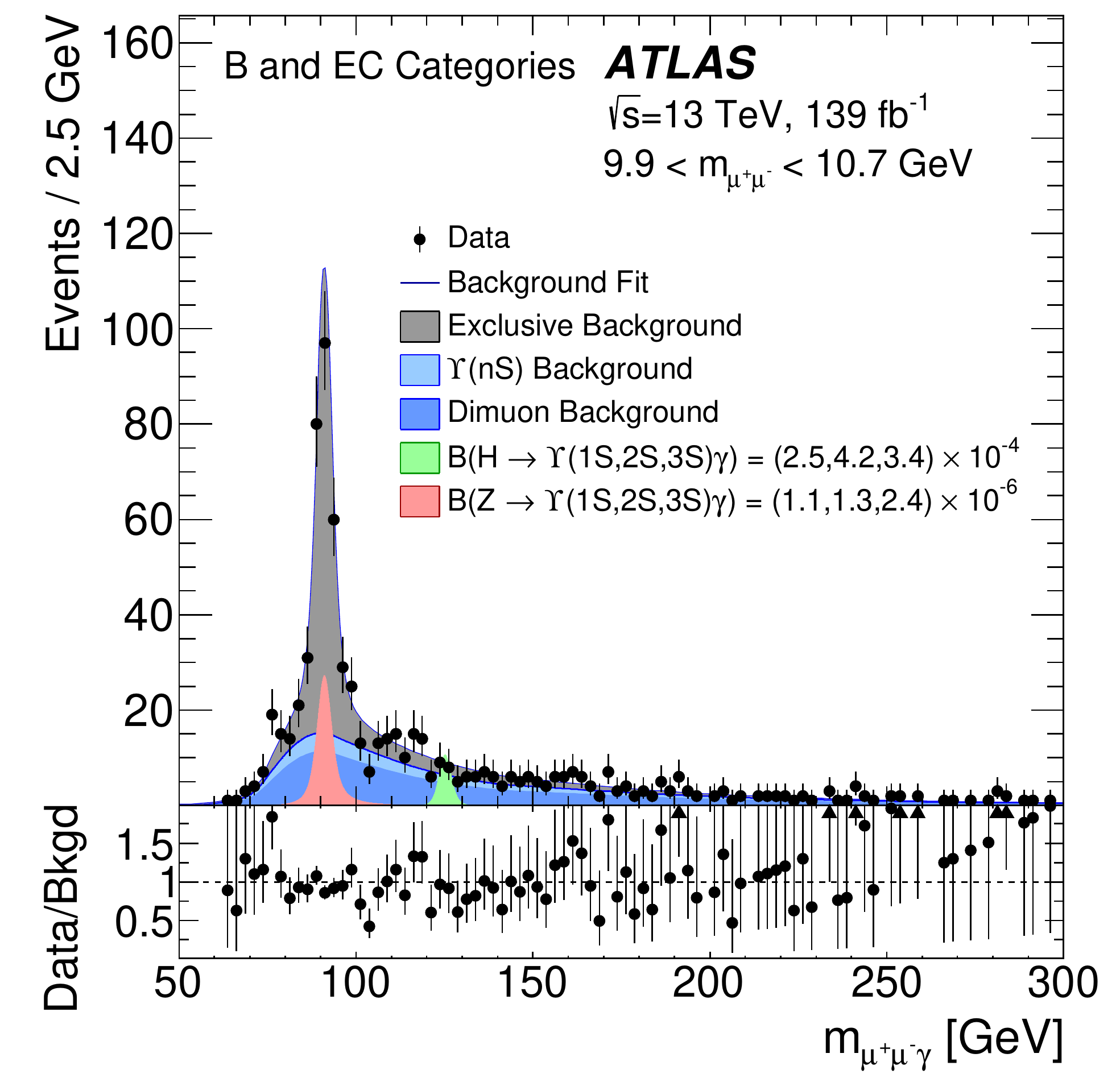}\label{fig:upsi_mumugamma_3SCut}}
\caption{Projection of the $\Upsilon(nS)$ channel fit in $m_{\mumu}$ for the~\subref{fig:upsi_mumu_ZCut} $Z$ boson and~\subref{fig:upsi_mumu_HCut} Higgs boson $m_{\mumu\gamma}$ regions, and in $m_{\mumu\gamma}$ for the $\Upsilon(1S,2S,3S)$ $m_{\mumu}$ regions in \subref{fig:upsi_mumugamma_1SCut}, \subref{fig:upsi_mumugamma_2SCut} and \subref{fig:upsi_mumugamma_3SCut}, respectively. The dimuon/${\cal Q}(nS)$ backgrounds in the legend refer to the inclusive background contribution which is non-resonant/resonant in $m_\mumu$. The branching fraction of each of the signals is set to the observed $95\%$ CL upper limit.\label{fig:upsilonFits}}
\end{figure}
 
\begin{table}[!htbp]
\centering
\caption{Numbers of observed and expected background events for the $m_{\mumu\gamma}$ ranges of interest.
Each expected background and the corresponding uncertainty of its mean is obtained from a background-only fit to the data; the uncertainty does not take into account statistical fluctuations in each mass range.
Expected $Z$ and Higgs boson signal contributions, with their corresponding total systematic uncertainty, are shown for reference branching fractions of $10^{-6}$ and $10^{-3}$, respectively.
The ranges in $m_{\mumu}$ are centred around each quarkonium resonance, with a width driven by the resolution of the detector; in particular, the ranges for the $\Upsilon(nS)$ resonances are based on the resolution in the endcaps.
It is noted that the discrepancy between the observed and expected backgrounds for $m_{\mumu} = 9.0$--$9.8~\GeV$ in the endcaps was found to have a small impact on the observed limit for $Z\rightarrow\Upsilon(1S)\,\gamma$.\label{tab:events}}
\begin{tabular}{|lc|rr|rr|rr|c}
\hline
&  & \multicolumn{4}{c|}{Observed (expected) background}&\multicolumn{1}{c|}{$Z$ signal} & \multicolumn{1}{c|}{$H$ signal}\\ \cline{3-6}
Category & $m_{\mumu}$ range & \multicolumn{4}{c|}{$m_{\mumu\gamma}$ range [\GeV]} & \multicolumn{1}{c|}{for} & \multicolumn{1}{c|}{for} \\ \cline{3-6}
& [\GeV] & \multicolumn{2}{c|}{$86$--$96$} & \multicolumn{2}{c|}{$122$--$128$}  & \multicolumn{1}{c|}{${\cal B}=10^{-6}$} & \multicolumn{1}{c|}{${\cal B}=10^{-3}$} \\ \hline
Inclusive         & 2.9--3.3\hphantom{0}  & 198 & $(185.6 \pm 5.9)$ & 61 & $(59.1 \pm 1.6)$ & $49.3 \pm 2.4$ & $87.8 \pm 6.1$ \\
Inclusive         & 3.5--3.9\hphantom{0}  &  83 & $(\hphantom{0}82.5 \pm 4.0)$ & 21 & $(22.9 \pm 0.9)$ & $6.5  \pm 0.3$ & $11.8 \pm 0.8$ \\
Barrel            & 9.0--9.8\hphantom{0}  & 125 & $(125.3 \pm 4.7)$ & 12 & $(11.6 \pm 0.6)$ & $11.4 \pm 0.6$ & $20.2 \pm 1.4$ \\
Barrel            & 9.6--10.4             & 118 & $(121.9 \pm 4.6)$ & 14 & $(10.7 \pm 0.6)$ & $8.8  \pm 0.4$ & $15.3 \pm 1.1$ \\
Barrel            & 9.9--10.7             & 102 & $(119.9 \pm 4.5)$ & 11 & $(10.2 \pm 0.6)$ & $10.1 \pm 0.5$ & $17.4 \pm 1.2$ \\
Endcap           & 9.0--9.8\hphantom{0}  & 133 & $(162.9 \pm 5.7)$ & 16 & $(13.6 \pm 0.7)$ & $15.5 \pm 0.8$ & $20.5 \pm 1.4$ \\
Endcap           & 9.6--10.4             & 150 & $(157.1 \pm 5.6)$ & 11 & $(11.7 \pm 0.5)$ & $11.7 \pm 0.6$ & $15.8 \pm 1.1$ \\
Endcap           & 9.9--10.7             & 171 & $(156.7 \pm 5.8)$ & 7  & $(11.4 \pm 0.6)$ & $13.5 \pm 0.7$ & $17.6 \pm 1.2$ \\
\hline
\end{tabular}
\end{table}
 
From the fit to the observed data, the largest observed local excess is 1.9$\sigma$ in the search for $Z\to\Jpsi\,\gamma$, followed by a
0.8$\sigma$ excess in the search for $H\to\psi(2S)\,\gamma$.
The expected and observed 95\% CL upper limits on the branching fractions for Higgs and $Z$ boson decays into a quarkonium state and a photon
are presented in Table~\ref{fig:Limit}, along with the observed upper limits in terms of Higgs and $Z$ boson production cross section times branching fraction to a quarkonium state and a photon.
The expected sensitivity improves by a factor of approximately two relative to the previous ATLAS result presented in Ref.~\cite{HIGG-2016-23};
this is in line with what is expected from the increase in integrated luminosity for this search, as this analysis is limited primarily by statistical uncertainties.
The systematic uncertainties in the signal normalisation and background shape described respectively in
Sections~\ref{sec:signalmodel} and~\ref{sec:backgroundmodel} result in a 0.8\% increase of the expected 95\% CL upper limit on the
branching fraction of the $H\to\jpsi\,\gamma$ decay. For the $Z\to\jpsi\,\gamma$
decay the effect is larger, 4.2\%, mostly due to the systematic
uncertainty in the background shape. The increase is 0.1\% for $H\to\psi(2S)\,\gamma$, and 0.6\% for $Z\to\psi(2S)\,\gamma$.
Similar behaviour is observed in the $\Upsilon(nS)\,\gamma$ analysis, with
systematic uncertainties resulting in a 0.1\%--0.8\% deterioration in the
sensitivity to the $H\to\Upsilon(1S,2S,3S)\,\gamma$ decays and a 0.3\%--0.5\% deterioration
in the sensitivity to the $Z\to\Upsilon(1S,2S,3S)\,\gamma$ decays.
 
For the interpretation of these searches in terms of constraints on the charm- and bottom-quark Yukawa couplings the approach presented in Refs.~\cite{Perez:2015aoa,Koenig:2015pha} is employed:
The ratio of signal strength ($\mu$) measurements for the $H\to J/\psi\,\gamma$ and $H\to\gamma\gamma$ channels corresponds to the ratio of measurements of their production cross section times branching fraction $\sigma\times\cal{B}$, normalised to their respective SM expectations. This is, to a good approximation, equal to the ratio of the respective partial decay widths $\Gamma$ normalised to their SM expectation $\Gamma^\text{SM}$, since the dependence on the production mechanism and Higgs total width cancels out. The ratio $\kappa_c/\kappa_\gamma$ of the coupling modifiers, each of which is the ratio of the coupling to its SM value, for the charm-quark Yukawa coupling and the effective coupling of Higgs boson to photons can be estimated as:
\[
\frac{\mu_{H\to J/\psi\,\gamma}}{\mu_{H\to\gamma\gamma}}=
\frac{\sigma_H\mathcal{B}_{H\to J/\psi\,\gamma} / \sigma^\text{SM}_H\mathcal{B}^\text{SM}_{H\to J/\psi\,\gamma}}{\sigma_H\mathcal{B}_{H\to\gamma\gamma}/\sigma^\text{SM}_H\mathcal{B}^\text{SM}_{H\to\gamma\gamma}}
\approx\frac{\Gamma_{H\to J/\psi\,\gamma}/\Gamma^\text{SM}_{H\to J/\psi\,\gamma}}{\Gamma_{H\to\gamma\gamma}/\Gamma^\text{SM}_{H\to\gamma\gamma}}=\frac{|{\cal A}_\text{ind}+{\cal A}_\text{dir}\kappa_c/\kappa_\gamma|^2}{\Gamma^\text{SM}_{H\to J/\psi\,\gamma}}
\]
The indirect and direct amplitudes ${\cal A}$ for $H\to J/\psi\,\gamma$ and $H\to \Upsilon(nS)\,\gamma$ interfere destructively and are obtained from Ref.~\cite{Brambilla:2019fmu}.\footnote{The corresponding direct and indirect amplitude values for $H\to\psi(2S)\,\gamma$ could not be obtained from the literature.}
The signal strength for $H\to\gamma\gamma$ is obtained from Ref.~\cite{ATLAS:2020pvn}.
An observed 95\% CL interval of $(-133,175)$ is obtained for $\kappa_c/\kappa_\gamma$,
with the expected interval being $(-120, 161)$.
The interval is dominated by the statistical uncertainty of the $H\to\jpsi\,\gamma$ search.
The correlated components in the uncertainties of the two measurements were removed, but this had negligible impact.
The theoretical uncertainties of the amplitudes result in a widening of the obtained interval by approximately 8\%, mainly through the uncertainty in the real part of the direct amplitude.
Furthermore, the magnitude of the direct amplitude, which is sensitive to the charm-quark coupling to the Higgs boson, is significantly smaller in the most recent theory calculations~\cite{Bodwin:2017wdu,Brambilla:2019fmu} than in earlier ones~\cite{HBranching}, leading to much weaker constraints.
Very large values of $\kappa_c$ lead to tensions with other ATLAS~\cite{HIGG-2018-57} and CMS~\cite{CMS-HIG-17-031} measurements of Higgs boson couplings~\cite{Coyle:2019hvs}.
A similar relation can be written for the ratio $\kappa_b/\kappa_\gamma$, where $\kappa_b$ is the coupling modifier for the bottom-quark Yukawa coupling.
Combining the three $\Upsilon(nS)\,\gamma$ decays, and accounting for the $-21\%$ correlation between $\mu_{H\to\Upsilon(2S)\,\gamma}$ and $\mu_{H\to\Upsilon(3S)\,\gamma}$, a 95\% CL interval of
$(-37,40)$
is obtained for $\kappa_b/\kappa_\gamma$.
The corresponding expected 95\% CL interval is
$(-37,39)$.
The $\Upsilon(1S)\,\gamma$ decay contributes most of the sensitivity to $\kappa_b/\kappa_\gamma$ thanks to its indirect amplitude being  the largest amongst the $\Upsilon(nS)\,\gamma$ decays.
Similarly to the $\kappa_c/\kappa_\gamma$ case, the statistical uncertainty of the search for exclusive Higgs boson decays into $\Upsilon(nS)\,\gamma$ dominates the interval.
The theoretical uncertainties of the amplitudes enlarge the interval by 12\%.
 
\begin{table}[h!]
\centering
\caption{Expected, with the corresponding $\pm 1\sigma$ intervals, and observed 95\% CL branching fraction upper limits
for the Higgs and $Z$ boson decays into a quarkonium state and a photon. Standard Model production of the Higgs boson is assumed. The corresponding upper limits on the production cross section times branching fraction $\sigma\times\cal{B}$ are also shown.
\label{fig:Limit}}
\begin{tabular}{l|rc|rc|c|c}
\cline{2-7}
\multicolumn{1}{c}{} & \multicolumn{6}{c}{95\% CL upper limits} \\ \cline{2-7}
\multicolumn{1}{c}{} & \multicolumn{4}{c}{Branching fraction} &\multicolumn{2}{|c}{$\sigma\times\cal{B}$} \\ \hline
\multicolumn{1}{c|}{Decay}   & \multicolumn{2}{c|}{Higgs boson $[\tabscript{10}{-4}{}]$} & \multicolumn{2}{c|}{$Z$ boson $[\tabscript{10}{-6}{}]$}& Higgs boson [fb] & $Z$ boson [fb]\\ \cline{2-7}
\multicolumn{1}{c|}{channel} & Expected & Observed & Expected & Observed  & Observed & Observed\\ \hline
$J/\psi\,\gamma$            & $\tabscript{1.8}{+0.8}{-0.5}$ & $2.0$ & $\tabscript{0.7}{+0.3}{-0.2}$ & $1.2$& 11 & 69\\
$\psi(2S)\,\gamma$          & $\tabscript{8.1}{+3.6}{-2.3}$ & $10.5$~~ & $\tabscript{3.0}{+1.3}{-0.8}$ & $2.4$& 58 & 142~~ \\
$\Upsilon(1S)\,\gamma$      & $\tabscript{2.7}{+1.2}{-0.8}$ & $2.5$ & $\tabscript{1.6}{+0.6}{-0.4}$ & $1.1$ & 14& 62\\
$\Upsilon(2S)\,\gamma$      & $\tabscript{3.4}{+1.5}{-1.0}$ & $4.2$ & $\tabscript{2.1}{+0.8}{-0.6}$ & $1.3$ & 24& 74\\
$\Upsilon(3S)\,\gamma$      & $\tabscript{3.0}{+1.3}{-0.8}$ & $3.4$ & $\tabscript{1.9}{+0.8}{-0.5}$ & $2.4$ & 19& 143~~ \\ \hline
\end{tabular}
\end{table}
\FloatBarrier


\section{Summary}
\label{sec:summary}
Searches for the exclusive decays of Higgs and $Z$ bosons into a vector quarkonium state and a photon have been performed with a $\rts=13~\TeV$ $pp$ collision
data sample collected with the ATLAS detector at the LHC and
corresponding to an integrated luminosity of $139~\ifb$.
The observed data are compatible with the background expectations.
The 95\% CL upper limits obtained for the $\Jpsi\,\gamma$ final state are
${\cal B}\left(H\to\Jpsi\,\gamma\right)<2.0\times10^{-4}$
and ${\cal B}\left(Z\to\Jpsi\,\gamma\right) < 1.2\times10^{-6}$. The corresponding upper limits for
the $\psi(2S)\,\gamma$ final state are ${\cal B}\left(H\to\psi(2S)\,\gamma\right)<10.5\times10^{-4}$
and ${\cal B}\left(Z\to\psi(2S)\,\gamma\right) < 2.4\times10^{-6}$.
The 95\% CL upper limits
${\cal B}\left(H\to\Upsilon(nS)\,\gamma\right) < (2.5,4.2,3.4)\times10^{-4}$ and
${\cal B}\left(Z\to\Upsilon(nS)\,\gamma\right) < (1.1,1.3,2.4)\times10^{-6}$
are set for the
$\Upsilon(1S,2S,3S)\,\gamma$ final states.
These upper limits represent an improvement by a factor of approximately two relative
to the previous results from the ATLAS Collaboration using
$36.1~\ifb$ of $\sqrt{s}=13~\TeV$ $pp$ collision data.
Further, constraints are set on the ratio $\kappa_c/\kappa_\gamma$ and $\kappa_b/\kappa_\gamma$ of the Higgs boson coupling modifiers.
An observed 95\% CL interval of $(-133,175)$ is obtained for $\kappa_c/\kappa_\gamma$, and a 95\% CL interval of
$(-37,40)$ is obtained for $\kappa_b/\kappa_\gamma$.


\section*{Acknowledgements}


We thank CERN for the very successful operation of the LHC, as well as the
support staff from our institutions without whom ATLAS could not be
operated efficiently.
 
We acknowledge the support of
ANPCyT, Argentina;
YerPhI, Armenia;
ARC, Australia;
BMWFW and FWF, Austria;
ANAS, Azerbaijan;
CNPq and FAPESP, Brazil;
NSERC, NRC and CFI, Canada;
CERN;
ANID, Chile;
CAS, MOST and NSFC, China;
Minciencias, Colombia;
MEYS CR, Czech Republic;
DNRF and DNSRC, Denmark;
IN2P3-CNRS and CEA-DRF/IRFU, France;
SRNSFG, Georgia;
BMBF, HGF and MPG, Germany;
GSRI, Greece;
RGC and Hong Kong SAR, China;
ISF and Benoziyo Center, Israel;
INFN, Italy;
MEXT and JSPS, Japan;
CNRST, Morocco;
NWO, Netherlands;
RCN, Norway;
MEiN, Poland;
FCT, Portugal;
MNE/IFA, Romania;
MESTD, Serbia;
MSSR, Slovakia;
ARRS and MIZ\v{S}, Slovenia;
DSI/NRF, South Africa;
MICINN, Spain;
SRC and Wallenberg Foundation, Sweden;
SERI, SNSF and Cantons of Bern and Geneva, Switzerland;
MOST, Taiwan;
TENMAK, T\"urkiye;
STFC, United Kingdom;
DOE and NSF, United States of America.
In addition, individual groups and members have received support from
BCKDF, CANARIE, Compute Canada and CRC, Canada;
PRIMUS 21/SCI/017 and UNCE SCI/013, Czech Republic;
COST, ERC, ERDF, Horizon 2020 and Marie Sk{\l}odowska-Curie Actions, European Union;
Investissements d'Avenir Labex, Investissements d'Avenir Idex and ANR, France;
DFG and AvH Foundation, Germany;
Herakleitos, Thales and Aristeia programmes co-financed by EU-ESF and the Greek NSRF, Greece;
BSF-NSF and MINERVA, Israel;
Norwegian Financial Mechanism 2014-2021, Norway;
NCN and NAWA, Poland;
La Caixa Banking Foundation, CERCA Programme Generalitat de Catalunya and PROMETEO and GenT Programmes Generalitat Valenciana, Spain;
G\"{o}ran Gustafssons Stiftelse, Sweden;
The Royal Society and Leverhulme Trust, United Kingdom.
 
The crucial computing support from all WLCG partners is acknowledged gratefully, in particular from CERN, the ATLAS Tier-1 facilities at TRIUMF (Canada), NDGF (Denmark, Norway, Sweden), CC-IN2P3 (France), KIT/GridKA (Germany), INFN-CNAF (Italy), NL-T1 (Netherlands), PIC (Spain), ASGC (Taiwan), RAL (UK) and BNL (USA), the Tier-2 facilities worldwide and large non-WLCG resource providers. Major contributors of computing resources are listed in Ref.~\cite{ATL-SOFT-PUB-2021-003}.


\printbibliography

\clearpage
\input{atlas_authlist}

\end{document}

%% file: atlas_authlist.tex
 
\begin{flushleft}
\hypersetup{urlcolor=black}
{\Large The ATLAS Collaboration}

\bigskip

\AtlasOrcid[0000-0002-6665-4934]{G.~Aad}$^\textrm{\scriptsize 101}$,
\AtlasOrcid[0000-0002-5888-2734]{B.~Abbott}$^\textrm{\scriptsize 119}$,
\AtlasOrcid[0000-0002-7248-3203]{D.C.~Abbott}$^\textrm{\scriptsize 102}$,
\AtlasOrcid[0000-0002-1002-1652]{K.~Abeling}$^\textrm{\scriptsize 55}$,
\AtlasOrcid[0000-0002-8496-9294]{S.H.~Abidi}$^\textrm{\scriptsize 29}$,
\AtlasOrcid[0000-0002-9987-2292]{A.~Aboulhorma}$^\textrm{\scriptsize 35e}$,
\AtlasOrcid[0000-0001-5329-6640]{H.~Abramowicz}$^\textrm{\scriptsize 150}$,
\AtlasOrcid[0000-0002-1599-2896]{H.~Abreu}$^\textrm{\scriptsize 149}$,
\AtlasOrcid[0000-0003-0403-3697]{Y.~Abulaiti}$^\textrm{\scriptsize 116}$,
\AtlasOrcid[0000-0003-0762-7204]{A.C.~Abusleme~Hoffman}$^\textrm{\scriptsize 136a}$
% ...
\bigskip
\\

% ...
$^{29}$Physics Department, Brookhaven National Laboratory, Upton NY; United States of America.\\
% ...
$^{35}$$^{(a)}$Facult\'e des Sciences Ain Chock, R\'eseau Universitaire de Physique des Hautes Energies - Universit\'e Hassan II, Casablanca;$^{(b)}$Facult\'{e} des Sciences, Universit\'{e} Ibn-Tofail, K\'{e}nitra;$^{(c)}$Facult\'e des Sciences Semlalia, Universit\'e Cadi Ayyad, LPHEA-Marrakech;$^{(d)}$LPMR, Facult\'e des Sciences, Universit\'e Mohamed Premier, Oujda;$^{(e)}$Facult\'e des sciences, Universit\'e Mohammed V, Rabat;$^{(f)}$Institute of Applied Physics, Mohammed VI Polytechnic University, Ben Guerir; Morocco.\\
% ...
$^{55}$II. Physikalisches Institut, Georg-August-Universit\"{a}t G\"ottingen, G\"ottingen; Germany.\\
% ...
$^{101}$CPPM, Aix-Marseille Universit\'e, CNRS/IN2P3, Marseille; France.\\
$^{102}$Department of Physics, University of Massachusetts, Amherst MA; United States of America.\\
% ...
$^{116}$Department of Physics, New York University, New York NY; United States of America.\\
% ...
$^{119}$Homer L. Dodge Department of Physics and Astronomy, University of Oklahoma, Norman OK; United States of America.\\
% ...
$^{136}$$^{(a)}$Departamento de F\'isica, Pontificia Universidad Cat\'olica de Chile, Santiago;$^{(b)}$Millennium Institute for Subatomic physics at high energy frontier (SAPHIR), Santiago;$^{(c)}$Instituto de Investigaci\'on Multidisciplinario en Ciencia y Tecnolog\'ia, y Departamento de F\'isica, Universidad de La Serena;$^{(d)}$Universidad Andres Bello, Department of Physics, Santiago;$^{(e)}$Instituto de Alta Investigaci\'on, Universidad de Tarapac\'a, Arica;$^{(f)}$Departamento de F\'isica, Universidad T\'ecnica Federico Santa Mar\'ia, Valpara\'iso; Chile.\\
% ...
$^{149}$Department of Physics, Technion, Israel Institute of Technology, Haifa; Israel.\\
$^{150}$Raymond and Beverly Sackler School of Physics and Astronomy, Tel Aviv University, Tel Aviv; Israel.\\
% ...
$^{a}$ Also Affiliated with an institute covered by a cooperation agreement with CERN.\\
$^{b}$ Also at An-Najah National University, Nablus; Palestine.\\
$^{c}$ Also at Borough of Manhattan Community College, City University of New York, New York NY; United States of America.\\
$^{d}$ Also at Bruno Kessler Foundation, Trento; Italy.\\
$^{e}$ Also at Center for High Energy Physics, Peking University; China.\\
$^{f}$ Also at Centro Studi e Ricerche Enrico Fermi; Italy.\\
% ...

\end{flushleft}
